\documentclass[manuscript,screen,nonacm]{acmart}
\makeatletter
\let\ACM@origbaselinestretch\baselinestretch
\makeatother
\usepackage{amsmath}
\usepackage{algorithmic}
\usepackage{graphicx}
\usepackage{textcomp}
\usepackage{xcolor}
\usepackage{capt-of}
\usepackage{minted}
\usepackage[ruled,linesnumbered]{algorithm2e}
\usepackage{import}
\usepackage{tikz}
\usepackage{moresize}
\usepackage{enumerate}
\usepackage{listings}
\usepackage{booktabs}
\usepackage{multirow}
\usepackage{colortbl}
\usepackage[inline]{enumitem}
\usepackage{soul}
\setstcolor{red}
\usepackage{csquotes}
\usepackage{url}

\usepackage{color}

\usepackage[most]{tcolorbox}

\usetikzlibrary{shapes.misc,decorations.pathreplacing,calligraphy,tikzmark}
\tcbuselibrary{minted, skins}
\usepackage[htt]{hyphenat}

\definecolor{Green}{HTML}{3fc380}
\definecolor{anotherred}{HTML}{E74C3C}
\definecolor{lightblue}{rgb}{0.9,0.9,1}
\definecolor{lightgreen}{rgb}{0.9,1,0.7}
\definecolor{lightred}{rgb}{1,0.9,0.9}

\makeatletter
\newcommand*{\rom}[1]{\expandafter\@slowromancap\romannumeral #1@}
\makeatother

\setminted[java]{
frame=lines,
framesep=1.5mm,
baselinestretch=1.0,
fontsize=\small,
linenos=false,
breaklines=true,
showtabs=false,
showspaces=false,
tabsize=2,
numberblanklines=false,
breakanywhere=true,
numbersep=3pt
}

\setminted[text]{
frame=lines,
framesep=2mm,
baselinestretch=1.2,
fontsize=\scriptsize,
linenos=false,
breaklines=true,
showtabs=false,
showspaces=false,
tabsize=2,
numberblanklines=false,
breakanywhere=true,
}

\newtcolorbox{rqbox}{
    colback=black!15!white,
    colframe=white,
    colbacktitle=white,
    coltitle=black,
    boxrule=1pt,
  sharpish corners,
  boxrule = 0pt,
    left=0mm,
    right=0mm,
    bottom=0mm,
    top=0mm,
  enhanced,
}

\def\BibTeX{{\rm B\kern-.05em{\sc i\kern-.025em b}\kern-.08em
    T\kern-.1667em\lower.7ex\hbox{E}\kern-.125emX}}

\newcommand{\code}[1]{\texttt{\small#1}}

\newcommand{\revised}[1]{\textcolor{black}{#1}}

\newcommand{\approach}[0]{\textsc{LLMSuite} }

\newcommand{\approachNS}[0]{\textsc{LLMSuite}}
\newcommand{\cmj}[0]{\mbox{\textsc{CodaMosa-J}}}

\newcommand{\circled}[1]{\tikz[baseline=(char.base)]{
            \node[shape=circle,draw,inner sep=1pt] (char) {#1};}}
\newcommand*\colorcircled[2]{\tikz[baseline=(char.base)]{
            \node[#1,shape=circle,draw,inner sep=1pt,align=center] (char) {#2};}}
\definecolor{commentbgcolor}{rgb}{0.8, 0.9, 0.9} %

\newcommand*\circledbeta{%
  \tikz[baseline=(char.base)]{%
    \node[shape=circle,draw,inner sep=1pt] (char) {$\beta$};}%
}
\newcommand*\circledalpha{%
  \tikz[baseline=(char.base)]{%
    \node[shape=circle,draw,inner sep=1pt] (char) {$\alpha$};}%
}

\definecolor{color1}{HTML}{FFE26F}
\definecolor{color2}{HTML}{0CA789}
\definecolor{color3}{HTML}{E74C3C}

\setcopyright{none}
\begin{document}

\title{Enhancing Automated Unit Test Generation for NLP Libraries Using Large Language Models}

 \author{Amirhossein Deljouyi}
 \email{a.deljouyi@tudelft.nl}
 \orcid{0009-0007-3405-8162}
 \affiliation{%
   \institution{Delft University of Technology}
   \city{Delft}
   \country{The Netherlands}
 }

 \author{Annibale Panichella}
 \email{a.panichella@tudelft.nl}
 \orcid{0000-0002-7395-3588}
 \affiliation{%
  \institution{Delft University of Technology}
   \city{Delft}
   \country{The Netherlands}}

 \author{Andy Zaidman}
 \email{a.e.zaidman@tudelft.nl}
 \orcid{0000-0003-2413-3935}
 \affiliation{%
  \institution{Delft University of Technology}
   \city{Delft}
   \country{The Netherlands}}

\renewcommand{\shortauthors}{Deljouyi et al.}

\begin{abstract}
Automated unit test generation tools like EvoSuite perform well on general-purpose software but often struggle with domain-specific software such as Natural Language Processing (NLP) libraries, where inputs must follow semantic, syntactic, and structural constraints. Large Language Models (LLMs) can generate domain-relevant test code, but tests produced by LLMs alone often fail to compile or achieve sufficient coverage.
We propose \textsc{LLMSuite}, a hybrid test generation framework that integrates self-refinement prompting with class-level LLM reasoning into the search-based testing process. In this mechanism, the LLM iteratively improves its test snippets based on feedback from previous generations. This enables the model to produce increasingly precise, domain-consistent code fragments that steer the evolutionary search toward exercising complex and otherwise hard-to-reach behaviors. When no objective improves over multiple generations in the underlying evolutionary algorithm, these refined snippets are parsed and injected into EvoSuite’s population to expand the search space.

To support our evaluation, we constructed a new dataset comprising 100 classes drawn from five widely used Java NLP projects. We also re-implemented \textsc{CodaMOSA}, a recent hybrid SBST–LLM technique, in Java to enable a direct comparison. Across this dataset, \textsc{LLMSuite} improves branch and line coverage by approximately 10\% and 8\%, respectively, and achieves an 11\% higher mutation score than \textsc{\cmj{}}. Compared to EvoSuite, \textsc{LLMSuite} yields roughly 15\% higher branch and line coverage and 5\% higher mutation score. Against an LLM-only baseline, it improves structural coverage by 36\% and mutation score by about 24.7 percentage points. Finally, \textsc{LLMSuite} complements manually written test suites by exercising domain-specific behaviors that are often left untested.
\end{abstract}



\maketitle

\section{Introduction}
\label{sec:intro}

In today’s software-driven world, ensuring software reliability and correctness is critical~\cite{KoCHASE2014,anicheSIGCSE2019}. As a result, automated (unit) testing has become a fundamental practice for software engineers aiming to deliver high-quality software~\cite{beck2003test-driven,khatamiSPE2024,khatamiSCAM2023}.
However, writing tests is a tedious and time-consuming task~\cite{beller2019developer,
aniche2022how-developers}. To reduce this burden, a variety of automated test generation techniques have been proposed~\cite{ali2010a-systematic,baresi2010testful:,fraser2011evosuite,fraser2015does,derakhshanfarTSE2023,brandtTSE2024}.
Notable tools in this area include Randoop~\cite{Pacheco_2007} and EvoSuite~\cite{fraser2011evosuite}. EvoSuite, for instance, is a search-based software testing (SBST) tool that leverages evolutionary algorithms to create test suites~\cite{FraserEMSE2015}, and has demonstrated strong performance in terms of code coverage~\cite{Fraser2013IEEE, Panichella2018AutomatedTC}, detecting software bugs~\cite{Shamshiri_2015, fraser20151600}, and helping developers during debugging~\cite{panichella2016impact}.

Although these tools are effective for general-purpose software, their applicability to specialized domains like natural language processing (NLP) and machine learning (ML) libraries remains largely unexplored~\cite{wang2021automatic}. This gap is especially concerning given the growing use of ML components in safety-critical systems, such as autonomous vehicles~\cite{gambi2019generating} and legal document pipelines~\cite{adedjouma2014automated}. NLP libraries are central to tasks like named entity recognition, sentiment analysis, and text classification, and are embedded in widely used frameworks such as Hugging Face Transformers. These libraries differ significantly from traditional software in structure, input expectations, and behavior~\cite{pauzi2023applications}. As a result, existing test generation tools often fail to produce effective test cases in these settings~\cite{wang2021automatic}.

\textit{Motivating Example.} Consider the \texttt{MorphaAnnotator} class from Stanford's CoreNLP\footnote{\url{https://github.com/stanfordnlp/CoreNLP/blob/v4.5.7/src/edu/stanford/nlp/pipeline/MorphaAnnotator.java}} in Listing~\ref{listing:motivating}, which processes phrasal verbs such as \texttt{gave\_up}. These cases require specific annotations that distinguish verb components. EvoSuite fails to generate test cases due to the:
\begin{enumerate*}
\item difficulty in constructing required objects, 
\item lack of awareness of domain-specific input formats, and 
\item inability to combine multiple relevant input properties.
\end{enumerate*}
Hence, approaches sensitive to domain-specific constraints are needed for such scenario.

Large Language Models (LLMs) have shown strong generative capabilities across both code and natural language~\cite{10329992, yu2023llm, mastropaolo2021icse, Liventsev_2023, wang2023software}. However, their ability to generate high-coverage, compilable unit tests for complex systems remains limited~\cite{elhajiAST2024, siddiq2024using, abdullin2025test}. In contrast, SBST tools excel at systematically exploring execution paths, but lack semantic and contextual understanding~\cite{abdullin2025test}.
Together, these limitations reveal an opportunity for a complementary hybrid approach.

The goal of our study is to understand where and how LLM-generated snippets contribute most effectively in the context of NLP libraries, whose inputs must satisfy strict linguistic and structural invariants (e.g., grammatically well-formed sentences, valid parse trees, and coherent annotation pipelines).

We propose \textsc{LLMSuite}, a hybrid framework that integrates LLMs into the SBST pipeline. The key idea is to leverage the complementary strengths of both approaches: LLMs provide contextual insights and semantically meaningful inputs, while SBST ensures systematic exploration through evolutionary search. \textsc{LLMSuite} employs class-level prompting combined with a self-refinement strategy, in which the LLM iteratively improves its outputs based on feedback from previous generations. We hypothesize that this design enables the LLM to generate increasingly precise, domain-consistent test snippets that guide SBST toward exercising complex and otherwise hard-to-reach behaviors.

Our study is guided by the following research questions:

\begin{enumerate}
\item[\textbf{RQ$_{1}$}] \textit{How does \approach compare to \cmj{}, standalone SBST and LLM-based methods in terms of code coverage and mutation score in NLP libraries?}
\end{enumerate}

\noindent RQ1 investigates whether combining LLMs with SBST leads to more effective test generation than either technique alone. To enable a fair comparison with a recent hybrid LLM–SBST approach, we reimplemented CodaMosa in Java and evaluated it alongside standalone SBST and LLM-based baselines. This RQ examines whether integrating LLM-generated snippets into the search process improves structural coverage and mutation score--areas where traditional tools often struggle in the context of NLP and machine learning libraries~\cite{wang2021automatic}.%

\begin{enumerate}
\item[\textbf{RQ$_{2}$}] \textit{How do the individual components of \approachNS~
influence the coverage of generated tests for NLP libraries?}
\end{enumerate}
\noindent RQ2 analyzes how \approachNS's design choices, e.g., the prompting strategy, choice of LLM model, refinement of string inputs, frequency of LLM test-generation calls, and using different context levels (method vs. class), affect coverage. 

\begin{enumerate}
    \item[\textbf{RQ$_{3}$}] \textit{How do SBST, LLM-based tests, and \approach perform across different functional categories of NLP libraries, and in what ways can \approach complement SBST?}
\end{enumerate}

\noindent NLP libraries implement a diverse set of functionalities ---such as tokenization, part-of-speech tagging, etc.--- each with distinct input constraints and structural characteristics. In RQ3 we examine whether LLMs can assist SBST in overcoming domain-specific challenges and local optima by generating more semantically relevant and diverse test inputs tailored to specific NLP tasks. We analyze  the effectiveness of each approach  across these categories.

\begin{enumerate}
    \item[\textbf{RQ$_{4}$}] \textit{To what extent can \approach complement manually written test cases in NLP libraries?}
\end{enumerate}

\noindent RQ4 explores whether \approachNS-generated tests cover scenarios not exercised by human-written tests.

\vspace{3pt}

\noindent The key contributions of this paper are as follows:
\begin{itemize}
    \item We propose \approach{}, a multi-mode hybrid test generation framework that integrates LLMs with SBST for domain-specific software testing.
    \item We conduct the first systematic evaluation of hybrid SBST/LLM test generation approach in the context of domain-specific libraries, focusing on NLP software.
    \item We construct a benchmark of 100 classes drawn from five widely used NLP libraries in Java to support evaluation of test generators in this setting.
    \item We release a public replication package with implementation, benchmarks, and results~\cite{replicationpackage}. 
\end{itemize}

\begin{listing}[!t]
\begin{minted}[escapeinside=??, fontsize=\footnotesize,linenos, 
 xleftmargin=20pt, numbersep=10pt]{java}
// A. a part of the method under test
private static String phrasalVerb(Morphology morpha, String word, String tag) {
  // must be a verb and contain an underscore
  if(!tag.startsWith("VB")  || !word.contains("_")) return null;
  // check whether the last part is a particle
  String[] verb = word.split("_");
  if(verb.length != 2) return null;
  String particle = verb[1];
  if(particles.contains(particle)) {
    String base = verb[0];
    String lemma = morpha.lemma(base, tag);
    return lemma + '_' + particle;
  }
  return null;
}
-------------------------------------------------------------------------------
// B. ChatGPT Testcase
@Test
public void testPhrasalVerbLemmatization() {
  MorphaAnnotator annotator = new MorphaAnnotator(false);
  CoreLabel token = new CoreLabel();
  token.set(TextAnnotation.class, "gave_up");
  token.set(PartOfSpeechAnnotation.class, "VBD");
  ...  
  annotator.annotate(annotation);
  String lemma = token.get(LemmaAnnotation.class);
  assertEquals("give_up", lemma);
}
\end{minted}
\caption{Motivating Example}
\label{listing:motivating}
\end{listing}

\section{Background}

\noindent\textbf{Search-Based Software Testing.~}
Automated test generation tools like EvoSuite~\cite{fraser2011evosuite} and Randoop~\cite{Pacheco_2007} generate test suites from Java code using search-based or random strategies~\cite{Fraser2013IEEE, Panichella2018AutomatedTC}. Search-Based Software Testing (SBST) formulates test generation as an optimization problem, using meta-heuristic algorithms guided by search objectives to maximize code coverage~\cite{Panichella2018AutomatedTC, rojas2016}. Seeding techniques, which inject prior knowledge (e.g., hard-coded constants strings from the class under test), can further improve effectiveness~\cite{rojas2016}. While Randoop uses feedback-directed random testing and scales well~\cite{wang2021automatic}, SBST generally achieves higher coverage, especially for hard-to-reach code~\cite{Panichella2018AutomatedTC}.

\vspace{2pt}
\noindent\textbf{Natural Language Processing Libraries.~}
Natural Language Processing (NLP), a branch of Artificial Intelligence (AI), enables computers to understand and generate human language by converting text into structured data~\cite{collobert2011naturallanguageprocessingalmost, chowdhary2020natural, hirschberg2015advances}. NLP has evolved from rule-based systems to statistical models, and more recently to machine learning (ML) and deep learning (DL), driven by advances in computing power and data availability~\cite{devlin2019bertpretrainingdeepbidirectional, zhao2022classificationnaturallanguageprocessing}. Core tasks include part-of-speech tagging, named entity recognition, machine translation, and question answering~\cite{chowdhary2020natural, devlin2019bertpretrainingdeepbidirectional}.
Libraries like \emph{Stanford CoreNLP} and \emph{NLTK}  provide tools and APIs for these tasks, often using embeddings or contextual models~\cite{manning2014stanford, bird2009natural}. 

\vspace{2pt}
\noindent\textbf{Large Language Models.~}
Large Language Models (LLMs) are AI systems based on the transformer architecture~\cite{vaswani2023attention}, trained on vast datasets to learn patterns in text, code, and dialogue~\cite{openai2023gpt4}. They generate responses through autoregressive token prediction. %
Prompt design plays a critical role in LLM effectiveness. Techniques like Chain of Thought (CoT) reasoning improve performance~\cite{wei2023chainofthought, li2023structured, marvin2023prompt},
Still, prompt construction often remains an empirical challenge~\cite{zamfirescu2023johnny}.

LLMs have been increasingly applied in software engineering~\cite{izadi2022codefill, izadi2024language, al2023extending}, with models like Code Llama~\cite{rozière2023code}, StarCoder~\cite{li2023starcoder}, Codex~\cite{DBLP:journals/corr/abs-2107-03374}, and GPT-4~\cite{openai2023gpt4} tailored for code-related tasks. Recent work explores their role in unit test generation~\cite{10329992, siddiq2024using, 10298372}, either as support in search-based testing~\cite{lemieux2023icse} or full automation~\cite{10329992}. Gu et al. have recently combined LLM-based test generation with static control flow analysis to address under-tested areas~\cite{guICSE2025}.

\section{The \approach Approach}\label{sec:approach}

\begin{figure}[!t]
\centering
\includegraphics[width=0.8\linewidth]{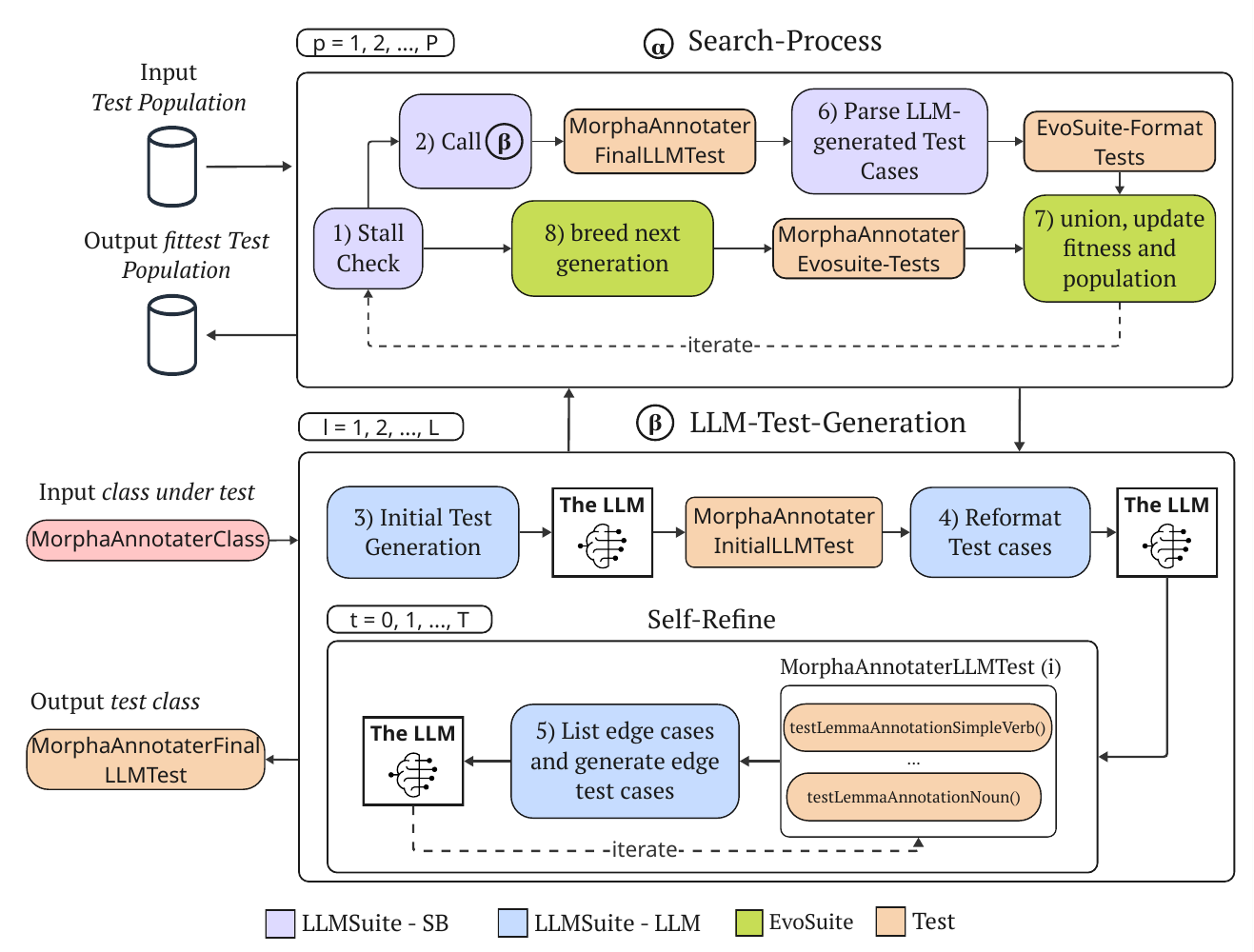}
\caption{Overview of the LLMSuite approach}
\label{fig_overview}
\end{figure}

Figure~\ref{fig_overview} presents an overview of our \approachNS~ approach. It consists of two main components: \circledalpha{} the Search Process and \circledbeta{} LLM-Based Test Generation.
\approach{} extends EvoSuite~\cite{fraser2011evosuite}, a search-based test generation framework, by integrating an LLM at key stages of the generation process. In particular, the LLM is invoked to generate diverse candidate test cases when the search process stagnates (highlighted in purple).
The integration is facilitated by additional components that ensure seamless interaction between EvoSuite and the LLM.
We also incorporate a self-refinement prompting strategy~\cite{madaan2023self} ---originally shown to improve performance by 20\% across seven tasks---, which we adapt to test generation to promote the generation of more diverse test cases (highlighted in blue).

The objective of \approachNS~ is to generate test cases that are either challenging or time-intensive to produce using conventional search-based techniques such as DynaMOSA~\cite{Panichella2018AutomatedTC}. 
\approachNS~ test generation is expected to provide advantages in scenarios involving:
\begin{enumerate*}
\item constructing complex test setups,
\item targeting edge cases and potential failure points, and 
\item generating test cases that require domain-specific knowledge, such as NLP libraries.
\end{enumerate*}

In our approach, when the fitness of the population does not improve for a certain number of generations ($S = 20$), we consider this a stagnation point (as~in~\circled{1}). At this stage, we trigger the LLM-Test-Generation process to introduce new test snippets
(\circled{2}).
These snippets are expected to bring in domain-specific knowledge that helps diversify the test pool and guide the search out of local optima.
To promote variety, the LLM is prompted multiple times with different goals. First, the LLM-Test Generation component generates the foundation of a test case (\circled{3}), then refactors it into the desired format (\circled{4}), and finally adds edge cases to target uncovered or hard-to-reach scenarios, resulting in more diverse and rich test cases in terms of coverage (\circled{5}).
These generated test cases are then parsed into a format compatible with search-based test generation (\circled{6}) and merged with the existing population (\circled{7}). In the following iterations, the search process can make use of these new, LLM-generated test cases ---which now embed domain knowledge--- to evolve additional tests that eventually reach new branches or exercise edge behaviors (\circled{8}).
Once the search completes, all test cases are compiled to ensure they are compilable and stable, meaning they should only fail due to assertion violations and not other issues (e.g., due to hallucinated code).

We now explain the LLM-test generation  and search process.

\subsection{LLM-Test Generation}\label{sec:approach:llm}

We use the \code{GPTx-4o} model from OpenAI~\cite{openai2023gpt4} as the default LLM in the test generation component of \approachNS. However, \approach{} is designed with a modular architecture, 
allowing it to easily replace the LLM with any other model. In Section~\ref{sec:result:llama}, we explore the effect of replacing ChatGPT with an alternative model, specifically, the open source \texttt{code-llama:13b-instruct} from Meta~\cite{rozière2023code}, accessed via the Ollama framework\footnote{Ollama: \url{https://ollama.com/}}.

\approach prompts the LLM in three sequential stages:
\begin{enumerate*}
\item generating the initial test class,
\item converting the generated test cases into the desired format, and
\item refining the tests to capture additional edge cases.
\end{enumerate*}
For each stage, we crafted specialized prompts based on best practices from recent prompt engineering research~\cite{marvin2023prompt, wei2023chainofthought, li2023structured, deljouyiICSE2025}. 
As illustrated in Listing~\ref{listing:enhanced_prompt}, these guidelines emphasize: adopting a clear developer persona (\circled{1}), using actionable and precise task instructions (\circled{2}), enabling deeper reasoning through techniques such as Chain-of-Thought and tailored guidance for each task (\circled{3}), and standardizing the input/output formats(\circled{4}).

We intentionally separate the stages of initial test generation and refactoring. This is because EvoSuite supports a specific format for test cases, which must be self-contained --- excluding constructs such as loops, external method calls, or helper utilities. Our trials showed that separating these two stages works better in practice; enforcing formatting constraints in the initial test generation process often reduces the LLM’s creativity and leads to guideline violations.
After receiving a response from the LLM, we apply post-processing steps, including syntax validation using ANTLR and automatic repair of common issues such as missing commas or brackets. \revised{We note that these repairs are performed at the LLM-based test generation stage, before any integration with EvoSuite, and should not be confused with the subsequent search process.}

\begin{figure}[!t]
\centering
\includegraphics[width=0.7\linewidth]{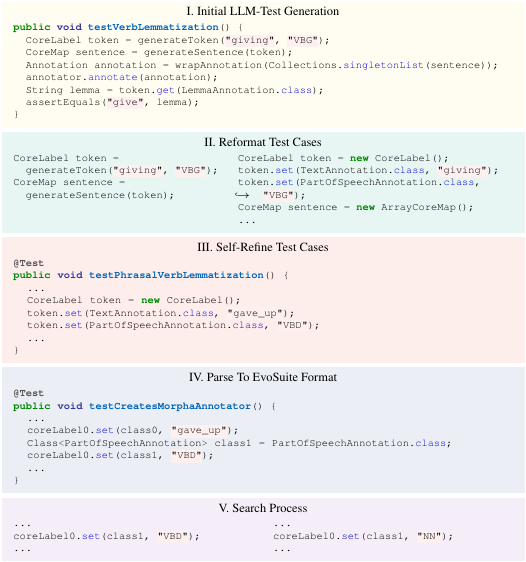}
\caption{Running Example}\label{fig:running_example}
\end{figure}

The LLM-Test-Generation component begins at step \circled{2} in Figure~\ref{fig_overview}, where it takes a class or method under test and generates corresponding test cases. By default, \approachNS~ uses the self-refinement prompting strategy with class-level prompting, which gives the LLM more context.
Figure~\ref{fig:running_example} illustrates this process using the \texttt{MorphaAnnotator} class: in Step~\rom{1}, the LLM produces an initial test case for lemmatizing the verb ``giving'', tagged as VBG. This test uses two external helper methods --- \texttt{generateToken} and \texttt{generateSentence} --- which need to be inlined into the test code during step~\circled{4} to match the format compatible with EvoSuite.
In step~\circled{5}, we initiate a self-refinement loop that runs for $T$ iterations (default $T=5$). In each iteration $t$, the LLM is prompted using a prompt template similar to Listing~\ref{listing:enhanced_prompt}. It is asked to analyze which scenarios have not yet been covered and then generate new edge test cases to fill those gaps. To enable this feedback loop, we maintain the conversational history and include it in the prompt so the LLM can reason over past test cases. All generated tests are added to a shared pool, and only unique cases --- based on their test names --- are retained. %
Unlike Madaan et al.'s work~\cite{madaan2023self}, which separates the feedback and refinement steps into two separate prompts, we unify them into a single prompt in order to reduce the number of prompts. In Step~\rom{3} of Figure~\ref{fig:running_example}, a test case for the phrasal verb case (``give\_up'') is generated as the initial test case.

\setcounter{lstlisting}{2}
\begin{listing}[!t]
\setlength{\parskip}{0pt}
\begin{minted}[escapeinside=??, breaklines, fontsize=\footnotesize]{text}
    ?\tikzmark{1:s}?<<SYS>>
        You are a (ADJECTIVE) developer focusing on (TASK AT HAND)
    ?\tikzmark{1:e}?<</SYS>>
    [INST]
?\footnotesize\colorcircled{blue}{2}?  Your task is to (TASK) while strictly following the guidelines below:
    ?\tikzmark{3:s}?### Guidelines:
      [For self-refinement:]
      1. Examine the existing test cases to identify untested scenarios.
      2. Clearly list these untested cases.
      3. Generate additional test cases to improve coverage, especially targeting
    ?\tikzmark{3:e}?     boundary conditions, corner cases, and potential failure points.
    ?\tikzmark{4:s}?### Output format:
      Your response should be placed between [TEST] and [/TEST].
      ### Code to Test:
    ?\tikzmark{4:e}?  [CODE]{content}[/CODE]
\end{minted}
\caption{Prompt template for LLMSuite}
\label{listing:enhanced_prompt}
\end{listing}
\begin{tikzpicture}[remember picture,overlay]

\draw[orange, line width=1pt, decorate, decoration={brace,
        raise=3pt, mirror,
        aspect=0.5}] (pic cs:1:s)  |- (pic cs:1:e) node[pos=0.25, left=2.5mm,inner sep=0em] {\footnotesize\circled{1}};

\draw[Green, line width=1pt, decorate, decoration={brace, mirror, raise=2pt}] 
  (pic cs:3:s) -- (pic cs:3:e) node[pos=0.50, left=2.5mm,inner sep=0em] {\footnotesize\circled{3}};

\draw[purple, line width=1pt, decorate, decoration={brace, mirror, raise=2pt}] 
  (pic cs:4:s) -- (pic cs:4:e) node[pos=0.50, left=2.5mm,inner sep=0em] {\footnotesize\circled{4}};
\end{tikzpicture}

\subsection{Search-Process}

EvoSuite generates test cases using a search-based approach guided by genetic algorithms. Among its algorithms, DynaMOSA has shown strong performance~\cite{Panichella2018AutomatedTC, CAMPOS2018207, lukasczyk2023empirical}. However, the search can stagnate in local optima --- particularly when test inputs require domain-specific knowledge, as seen in our motivating example.
To address this, we extend DynaMOSA by using LLM-generated tests to get unstuck from stagnation: When no improvement is observed across any objective for 20 consecutive generations—a scenario DynaMOSA identifies using its many-objective optimization mechanism—
\approach invokes an LLM to generate new test cases.
These LLM-generated tests inject domain semantics into the test population --- semantic knowledge that is difficult for the search alone to infer. This phase represents a key moment where domain semantics meet search. %
While inspired by prior work~\cite{lemieux2023icse}, our approach differs substantially in both design and execution. Unlike CodaMOSA’s one-shot queries without advanced prompt engineering, \approachNS~ guides the LLM through an iterative self-refinement process, where candidate tests are revised using feedback about uncovered scenarios; this search-like mechanism—drawing on the self-refine paradigm~\cite{madaan2023self}—enables the generation of more complex, domain-consistent inputs that SBST can exploit more effectively. \approachNS~ also adopts class-level prompting, in contrast to CodaMOSA’s method-level design, reducing the number of prompts while providing richer context for reasoning about interactions across methods. In addition, \approachNS~ integrates with DynaMOSA and introduces a custom parsing and validation pipeline that supports nested invocations, varargs, multi-dimensional arrays, assignments, and mock statements, thereby allowing \approach{} to use most LLM-generated code snippets.

The generated test cases are parsed into EvoSuite compatible code by \emph{LLM2EvoSuiteParser}; our custom parser uses Spoon~\cite{spoon} for static analysis and converts the textual code generated by LLM into the internal representation of EvoSuite. For instance, a line like \texttt{token.set(..., "VBD")} becomes an object initialization and method call, such as \texttt{class1} and  \texttt{coreLabel0.set(...)}. The parser supports a wide range of constructs, including nested invocations, varargs, multi-dimensional arrays, and assignments.

When encountering unsupported constructs in EvoSuite during the search phase (e.g., loops, lambdas, try-catch blocks), the parser selectively extracts valid subcomponents, such as expressions in assertions or try-block statements. It also tries to skip the errors of hallucinated lines. For example, if token.set(...) has three parameters in the LLM-generated test snippet but only two are valid, the parser retains the first two and discards the rest by loose parameter matching.

After parsing, valid LLM-generated tests are unioned with the current population. EvoSuite ranks and updates the population before proceeding to the next generation. This LLM integration can occur up to L times (default L = 5), and LLM-generated tests can participate in genetic operations such as crossover and mutation. As a result, EvoSuite can explore test inputs it could not have evolved alone,
e.g., in Example~\rom{5} of Figure~\ref{fig:running_example}, a phrasal verb like ``give\_up'' is crossed with another test tagged ``NN'', producing a variant that triggers an exception on line 11 of Listing~\ref{listing:motivating}. These hybrid cases demonstrate the value of integrating LLM-generated content into search-based testing for broader coverage.

\vspace{-5pt}
\section{Experiment Setup}\label{sec:experiment}

In this section we describe the methodology of evaluation of our approach considering the RQs introduced in Section~\ref{sec:intro}.
\vspace{2pt}

\begin{enumerate}
\item[\textbf{RQ$_{1}$}] \textit{How does \approach compare to \cmj{}, standalone SBST, and LLM-based methods in terms of code coverage, mutation score in NLP libraries?}

\item[\textbf{RQ$_{2}$}] \textit{How do the indivual components of \approach{}
the individual components of our design 
influence the coverage of generated tests for NLP libraries?}
\item[\textbf{RQ$_{3}$}] \textit{How do \approach tests perform across different functional categories of NLP libraries, and in what ways can \approach complement SBST?}
\item[\textbf{RQ$_{4}$}] \textit{To what extent can \approach complement manually written test cases in NLP libraries?}
\end{enumerate}

\noindent\textbf{Baselines Selection}.
\textit{RQ1 and RQ3.} We consider three baselines for comparison in RQ1 and RQ3: 
\begin{enumerate*}
    \item \textit{EvoSuite}~\cite{fraser2011evosuite}, configured with the DynaMosa search strategy~\cite{Panichella2018AutomatedTC},  
    \item \textit{LLMOnly}, which consists of test cases generated solely by \revised{ChatGPT-4o-latest-2025-07-12},~\cite{openai2023gpt4}, and
    \item \textit{\cmj{}}~\cite{lemieux2023icse}, using our Java re-implementation of the approach. Since the original CodaMosa was designed for Python, we reimplemented the high-level algorithmic logic as closely as possible-- including the walkthrough procedure, the focus on low-coverage test targets, method level prompting, and the use of MOSA as the underlying SBST strategy. However, several components necessarily differ from the original implementation: CodaMosa’s study used OpenAI Codex~\cite{DBLP:journals/corr/abs-2107-03374}, while we use ChatGPT-4o-latest, and we rely on our own Java parser in place of the Python-specific tooling used originally.

\end{enumerate*}
\emph{LLMOnly} test cases are similar to those used in \approach but are not processed by our parser. As a result, this baseline may achieve higher raw coverage, due to the parser's limitations in supporting certain code constructs.

\textit{RQ2.} For RQ2, we conduct a systematic analysis of how individual components of our approach affect test coverage. Specifically, we examine:
\begin{enumerate*}
\item the choice of LLM model, as well as the level of contextual information provided (method vs.\ class),
\item the prompting strategy, including the use of self-refinement,
\item techniques for refining string inputs,
\item the frequency of invoking LLM-based test generation.
\end{enumerate*}
To isolate the effect of each factor, we design four experimental variants, each modifying a single component while holding others fixed. We evaluate test coverage of each variant in relation to both EvoSuite and the default \approach configuration.

\textit{RQ4.} 
In RQ4, we compare the test coverage achieved by \approachNS~ with that of manually written test cases. This comparison is limited to the subset of classes in the dataset for which manually written tests are available.

\noindent\textbf{Dataset and NLP Projects}.
We have collected 100 classes from five widely-used Java-based NLP libraries: CoreNLP~\cite{manning2014stanford},  OpenNLP~\cite{opennlp}, MALLET~\cite{mccallum2002mallet}, 
CogCompNLP~\cite{2018_lrec_cogcompnlp}, and
GATE~\cite{cunningham2002gate}.
Table~\ref{table:projects} summarizes key statistics for these libraries. We adapted our approach and tooling to be compatible with Java 8, 11, and 17.

\begin{table}[!t]
\centering
\caption{Overview of NLP Libraries Used in the Experiment}
\footnotesize
\begin{tabular}{l p{5.5cm} c c c}
\toprule
Project & Description & Version & Java & \#Classes\\
\midrule
CoreNLP~\cite{manning2014stanford} & Full-featured NLP toolkit with parsing, NER, and coref & 4.5.7 & 11 & 42 \\
OpenNLP~\cite{opennlp} & Basic NLP tasks like tokenization and POS tagging & 2.3.3 & 17 & 17 \\
Mallet~\cite{mccallum2002mallet} & Text classification and sequence tagging & 2.1.0 & 8 & 16 \\
GATE~\cite{2018_lrec_cogcompnlp} & Rule-based and ML-based NLP pipeline & 9.1.0 & 8 & 11 \\
CogCompNLP~\cite{2018_lrec_cogcompnlp} & Semantic and coreference analysis toolkit & 4.0.15 & 8 & 14 \\
\bottomrule
\end{tabular}
\label{table:projects}
\end{table}

\textit{Collecting Target Classes from NLP Libraries}.
To collect classes from the selected NLP libraries, we followed several steps (Filtering, Classifying \& Collecting, and Categorization)  to ensure a balanced and representative dataset. 

\begin{itemize}
 \item Filtering:
 \begin{enumerate*}
     \item We filter out all static, abstract, private classes from our search.
     \item For each remaining class, we computed three metrics:
     \begin{enumerate*}
       \item Weighted Methods per Class (WMC),
       \item Number of branches, and
       \item Number of non-static methods.
     \end{enumerate*}
     These metrics were normalized and combined into a single score.
     \item Based on this score, we ranked the classes and selected the top 50 classes for each project.
 \end{enumerate*}
 \item Classifying \& Collecting:
 \begin{enumerate*}
 \item We manually reviewed their JavaDoc documentation to assign a functional category, based on the categorization scheme from~\cite{2018_lrec_cogcompnlp}.
 \item From this set, we selected 100 classes that together represent a diverse range of functionalities.
 \end{enumerate*}
 \item Categorization: We grouped the functional categories into broader semantic categories for analysis.
\end{itemize}
Finally, we organized the dataset into nine functional categories, see Table~\ref{table:categories}.
To reduce the number of runs in RQ2, we randomly selected 60 of the 100 target classes.

\begin{table}[t]
\footnotesize
\setlength{\tabcolsep}{2pt}
\caption{Functional categories in NLP software}\label{table:categories}
\begin{tabular}{lp{12cm}}
\toprule
Coreference &  covers components that detect and resolve references to the same entity, like mention detection and coreference resolution\\
Data Structures & includes core data classes used to represent text, e.g., documents and corpora, and supporting utility functions.\\
Information Extraction &  components that extract specific information from text—such as quotes, names, gender, or relational features --- often rule-based without relying on external models.\\
Linguistic Labeling & tasks like assigning labels to text, such as NER, POS tagging, sentiment analysis, gender recognition, and wikification\\
ML Algorithms & learning algorithms like classifiers and clustering, often used to power labeling and coreference modules.\\
Normalization & covers standard text normalization processes like stemming and lemmatization.\\
Parser & components that analyze the syntactic structure of text --- such as dependency or constituency parsing --- and generate tree or graph representations that support downstream tasks.\\
Segmentation & involves breaking down text into units, e.g., tokenization, sentence splitting, and chunking.\\
Topic Modeling & focuses on uncovering latent topics in text using probabilistic models like LDA.\\
\bottomrule
\end{tabular}
\end{table}

\begin{table}[!t]
\centering
\caption{NLP Functional Category Distribution of Dataset}
\vspace{-3mm}
\setlength{\tabcolsep}{2pt}
\footnotesize
\begin{tabular}{lcccccc}
\toprule
Category & CoreNLP & OpenMLP & Mallet & Gate & CogCompNL & Total\\
\midrule
Coreference& 7 & 0 & 0 & 0 & 1 & 8\\
DataStructure \& Utils& 3& 0 & 2& 7 & 1 & 13\\
Information Extraction& 5& 1 & 0& 0 & 1 & 7\\
Linguistic Labeling& 9& 2 & 0& 0& 4 & 15\\
ML Algorithms& 0& 1 & 9& 1 & 3 & 14\\
Normalization& 3& 5 & 0& 0 & 1 & 9\\
Parser& 8& 2 & 0& 3 & 2 & 15\\
Segmentation& 7& 3 & 0& 0 & 1 & 11\\
Topic Modeling& 0& 3 & 5& 0 & 0 & 8\\
\midrule
Total& 42 & 17 & 16 & 11 & 14 & 100 \\
\bottomrule
\end{tabular}
\label{table:classes}
\end{table}

\noindent\textbf{Metrics for Evaluating Effectiveness}.
To assess the quality of a unit test suite for a given project, we use code coverage and mutation score.
For coverage, we use metrics inspired by EvoSuite, which include instruction and branch coverage. Instruction coverage corresponds to Java bytecode instructions and is roughly equivalent to statement coverage at the source code level. Branch coverage measures whether both outcomes (true and false) of each conditional statement have been executed, effectively covering the edges in the program’s control flow graph.
For mutation score, we utilize  JUGE~\cite{Devroey2022}, a benchmarking infrastructure for evaluating Java unit test generators and it has been widely used in the yearly SBFT tool competitions~\cite{DBLP:conf/icse/JahangirovaT23, moon2025evofuzz}. JUGE internally uses PITest~\cite{pitest} to compute mutation scores, which quantify the ability of a test suite to detect injected faults (mutants). We extended JUGE to support Java versions 11 and  17 to ensure compatibility with our benchmark.

\noindent\textbf{Analysis Method}.
To compare the test effectiveness of \approachNS~against the baseline tools across all target classes, we used a repeated-measures design, following established evaluation guidelines for search-based techniques\cite{arcuri2014hitchhiker} and studies involving LLMs~\cite{sallou2024breaking}. Each tool was executed five times per class to capture variability arising from inherent non-determinism in both LLM-based and SBST approaches.
For the LLM component, we used the default temperature of 1.00 (range [0,2]). Lower temperatures make outputs more deterministic, while higher ones introduce more randomness. Although prior work shows that temperature does not correlate with output quality~\cite{10.1145/3661167.3661221}, using the default value offers a consistent and reasonable level of stochasticity for evaluating LLM-based test generation.

To compare each tool on a per-class basis, we applied the Wilcoxon signed-rank test ($\alpha = 0.05$), which is suitable for paired, non-normally distributed data. To compare LLMSuite against the other baselines across all CUTs, we first computed, for each metric (branch coverage, line coverage, and mutation score), the median over the five runs as the representative performance value. We then used the Friedman test to assess whether the tools differ significantly in their median scores, followed by Conover’s post-hoc test for pairwise multiple comparisons.
To quantify the magnitude of differences, we computed Vargha and Delaney’s $A_{12}$ effect size~\cite{vargha2000critique}, which represents the probability that one method outperforms another. $A_{12}$ values are interpreted as small ($0.56 \le A_{12} < 0.64$), medium ($0.64 \le A_{12} < 0.71$), and large ($A_{12} \ge 0.71$) effects~\cite{vargha2000critique}. Finally, to assess the robustness of the effect-size estimates, we computed 95\% bias-corrected bootstrap confidence intervals (CI) over the per-class $A_{12}$ values, following Efron’s bootstrap technique~\cite{efron1992bootstrap}. A 95\% confidence interval provides an uncertainty range within which the true effect size would fall in 95\% of repeated samples, offering a distribution-free way to account for randomness inherent in both LLM-based and search-based test generation. For each class, we generated thousands of bootstrap resamples and recalculated $A_{12}$; the lowest and highest bootstrap estimates form the CI bounds. We determined per-class significance by checking whether the CI lies entirely above 0.5 (LLMSuite wins), entirely below 0.5 (baseline wins), or crosses 0.5 (tie).

\noindent\textbf{Experimental Protocol}.
We configured EvoSuite with its default parameters, which have shown good performance in prior work~\cite{arcuri2013parameter}. To allow more time for generating high-quality test cases, we increased the test budget (\texttt{max\_time}) from 60 to 900 seconds.
To ensure reliable statistical comparisons across the seven test generators --- EvoSuite, \cmj{}, \approachNS and its four variants --- we repeated each run five times, \revised{totaling 3300 runs ($(100 \times 3 + 60 \times 6) \times 5$)}, where 100 classes were tested with \approach{} and EvoSuite and 60 classes \revised{with 6 variants, each repeated 5 times}. We ran experiments on a server with a 64-core AMD EPYC processor and 256 GB RAM. We used Docker to parallelize execution, allocating 6 CPU cores and 16 GB RAM per container.
\approachNS~ uses ChatGPT-4o-latest-2025-07-12\footnote{\url{https://platform.openai.com/docs/models/chatgpt-4o-latest}} as its primary LLM (see Section~\ref{sec:approach:llm}). To evaluate the impact of model choice, we also ran \texttt{code-llama:13b-instruct}~\cite{rozière2023code} on an NVIDIA L40S GPU.

\noindent\textbf{Manual Analysis}.
For RQ3 and RQ4, we conducted a manual analysis of test coverage.
For RQ3, we examined classes with notable coverage differences between EvoSuite and \textsc{LLMSuite}, spanning various NLP functionalities. By identifying which parts of each class were covered by one approach but missed by the other, we gained insights into their relative strengths and limitations across different NLP tasks.
For RQ4, we analyzed 27 classes that had both manually written test cases and were part of our dataset. \revised{For these classes, we first calculated branch coverage and mutation score for manually written tests, \approachNS, and their combination.} We then identified code blocks left uncovered by both the manual tests and \approachNS, and investigated the underlying reasons for these gaps.
Since the dataset was not very large, involving multiple coders was unnecessary. Instead, we followed Hoda’s guidance for single-analyst studies~\cite[p.~260]{hoda2025qualitative}. The primary analyst conducted the coding, and the emerging codes and interpretations were regularly reviewed with the other researchers to ensure consistency and refine the analysis where needed.

\begin{figure*}[!t]
\centering
\includegraphics[width=\linewidth]{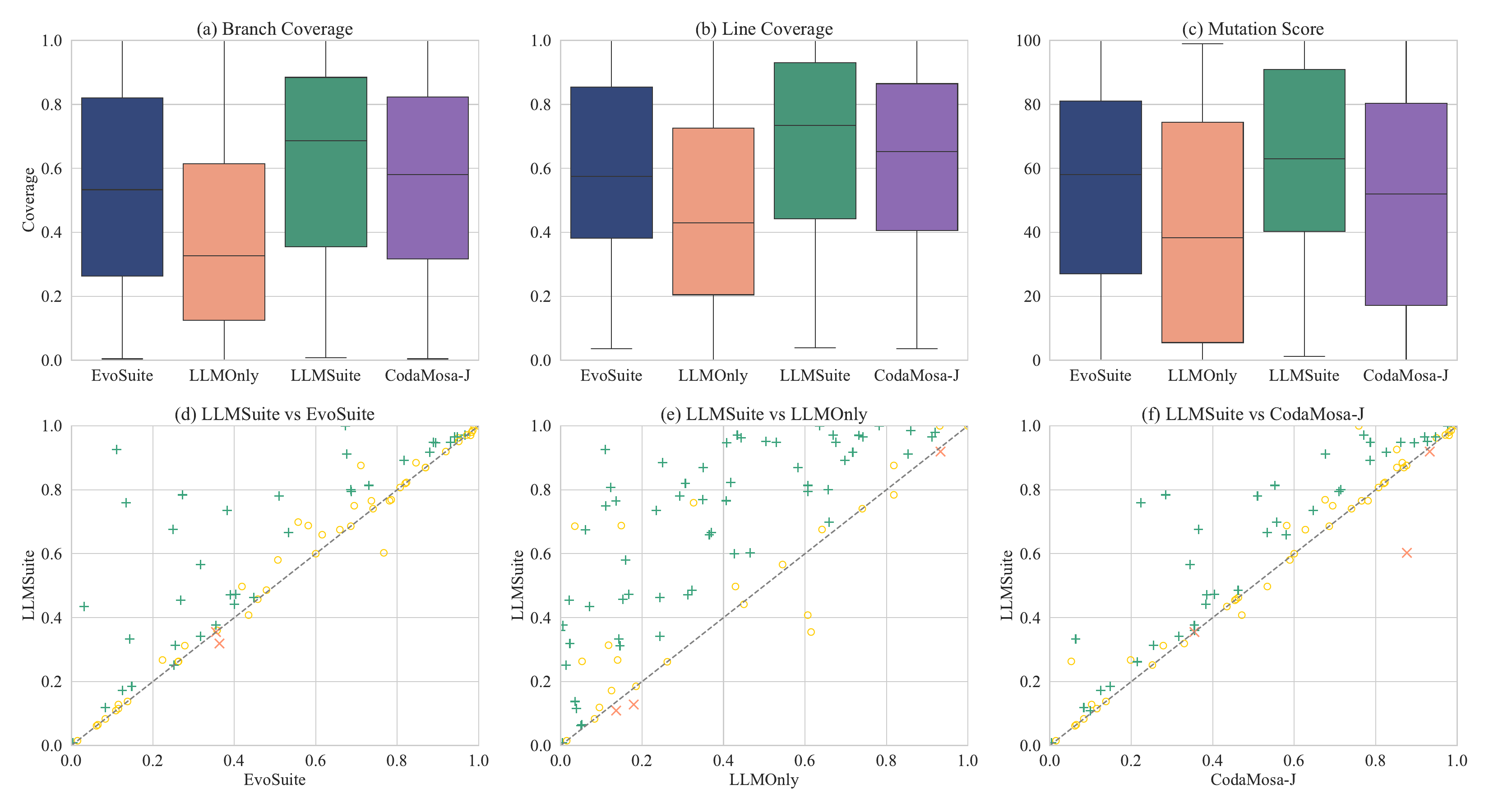}
\vspace{-6mm}
\caption{Comparison of test effectiveness across LLMSuite and the baseline tools.
(a) Branch coverage, (b) line coverage, and (c) mutation score distributions across all classes, shown as box plots.
(d–f) Scatter plots comparing per-class coverage between \approachNS{} (y-axis) and each baseline (x-axis): EvoSuite (d), LLMOnly (e), and \cmj{} (f). Points on the diagonal indicate identical coverage, points above the diagonal represent classes where \approachNS{} achieves higher coverage, and points below denote baseline wins.}\label{fig:results_all}. 
\vspace{-3mm}
\end{figure*}

\vspace{-5pt}
\section{Result}
In this section, we present and discuss the results for each research question.

\subsection{RQ1: Effectiveness of LLMSuite vs. Baselines}\label{sec:rq1}

Figure~\ref{fig:results_all} (a–b) shows box plots comparing line and branch coverage across the evaluated tools. \approachNS~ consistently outperforms all three \textsc{\cmj{}}, \textsc{EvoSuite} and \textsc{LLMOnly} in terms of median coverage. For branch coverage, \approachNS~ reaches a median of 68.60\%, versus 58.03\% for \textsc{\cmj{}} 53.33\% for \textsc{EvoSuite} and 32.64\% for \textsc{LLMOnly}. Line coverage follows a similar trend: \approachNS~ achieves 73.49\%, compared to 65.62\%, 57.66\% and 35.94\% for \textsc{\cmj{}}, \textsc{EvoSuite} and \textsc{LLMOnly}, respectively. These results correspond to improvements of 10.57\% in branch coverage and 7.87\% in line coverage over \textsc{\cmj{}}. Compared to \textsc{EvoSuite}, the gains are 15.27\% for branch coverage and 15.94\% for line coverage, on average.
Project-level results for branch coverage, line coverage, and mutation score are reported in Table~\ref{tab:projects_coverage}. The improvements are consistent across all projects, though the gains for \textit{CoreNLP} are particularly pronounced.

We used the Wilcoxon signed-rank test to evaluate statistical significance. The results show that LLMSuite-generated test cases achieve significantly higher branch and line coverage than those generated by \cmj{} and EvoSuite ($p\textrm{-value} \ll 0.05$), and also significantly outperform LLMOnly ($p\textrm{-value} \ll 0.05$). The exact values are reported in Table~\ref{tab:statistics}.
The $A_{12}$ effect sizes are large when comparing LLMSuite with \cmj{} (0.73) and EvoSuite (0.72), and even larger when compared to LLMOnly (0.85), with confidence intervals consistently above 0.5.
Since each tool was run five times per class, we performed per-class significance testing. 
Figure~\ref{fig:results_all}(d–f) shows the resulting win-rate patterns. The analysis reveals a clear trend: LLMSuite achieves higher branch coverage in 38 classes compared to \cmj{} (and lower in only 3), in 36 classes compared to EvoSuite (lower in just 2), and in 59 classes compared to LLMOnly (lower in only 3).

\begin{table}[b!]
\centering
\caption{Comparison of Median Branch Coverage, Median Line Coverage, and Median Mutation Score Across Projects}
\renewcommand{\tabcolsep}{4.5pt}
\begin{smaller}
\begin{tabular}{lc|cccc|cccc|cccc}
\toprule
\multirow{2}{*}{\textbf{Project}} &
\multirow{2}{*}{\textbf{\# }} & 
\multicolumn{4}{c}{\textbf{Branch Coverage}} &
\multicolumn{4}{c}{\textbf{Line Coverage}} &
\multicolumn{4}{c}{\textbf{Mutation Score}} \\
\cline{3-6}
\cline{7-10}
\cline{11-14}
& & \textbf{LLS} & \textbf{CM} & \textbf{ES} & \textbf{LLM} 
  & \textbf{LLS} & \textbf{CM} & \textbf{ES} & \textbf{LLM}
  & \textbf{LLS} & \textbf{CM} & \textbf{ES} & \textbf{LLM} \\
\midrule
CoreNLP & 42 
& 48.5\% & 33.69\% & 29.8\% & 16.01\%
& 61.62\% & 42.03\% & 41.66\% & 28.47\%
& 43.08\% & 17.39\% & 35.47\% & 33.25\% \\
OpenNLP & 17 
& 88.84\% & 82.47\% & 81.97\% & 47.7\%
& 92.37\% & 86.77\% & 86.58\% & 62.37\%
& 70.0\% & 68.0\% & 60.50\% & 39.5\% \\
Mallet & 16 
& 78.79\% & 72.84\% & 77.62\% & 44.97\%
& 84.03\% & 78.88\% & 82.12\% & 56.77\%
& 89.82\% & 81.69\% & 80.21\% & 26.37\% \\
Gate & 11 
& 53.32\% & 52.61\% & 49.3\% & 27.82\%
& 58.09\% & 57.20\% & 54.72\% & 25.87\%
& 73.0\% & 70.0\% & 72.5\% & 59.0\% \\
CogCompNLP & 14 
& 37.63\% & 35.48\% & 31.71\% & 17.95\%
& 44.27\% & 42.99\% & 43.31\% & 25.82\%
& 62.0\% & 52.0\% & 62.0\% & 38.0\% \\
\midrule
Overall & 100 
& \textbf{68.60\%} & 58.03\% & 53.33\% & 32.64\%
& \textbf{73.49\%} & 65.62\% & 57.50\% & 43.00\%
& \textbf{63.0\%} & 52.0\% & 58.0\% & 38.26\% \\
\bottomrule
\end{tabular}
\end{smaller}
\label{tab:projects_coverage}
\end{table}

In Figure~\ref{fig:results_all}(d), we observe that \textsc{LLMSuite} consistently outperforms \textsc{\cmj{}}, \textsc{EvoSuite}, and \textsc{LLMOnly} in terms of mutation score, although the effect sizes are more moderate than those seen for structural coverage. At the median, \approachNS~ achieves a mutation score of 63.0\%, compared to 52.0\% for \textsc{\cmj{}}, 58.0\% for \textsc{EvoSuite}, and 38.26\% for \textsc{LLMOnly}. This corresponds to improvements of 11\% over \cmj{}, 5\% over EvoSuite, and an approximately 24.7 percentage-point gain over LLMOnly.

These differences are statistically significant: the Wilcoxon signed-rank test reveals p-values of 0.0003 (vs. \cmj{}), 0.007 (vs. EvoSuite), and 0.002 (vs. LLMOnly). The $A_{12}$ effect sizes--0.58, 0.57, and 0.64, respectively--indicate consistent, if more modest, advantages, with all 95\% confidence intervals remaining above 0.5. Taken together, these results indicate that LLMSuite achieves higher mutant-kill rates than the baselines, although the magnitude of improvement is more modest compared to structural coverage metrics.

Despite these gains, we observe that \approachNS~ does not gain as much benefit in mutation score as it does in structural coverage. We attribute this to the fact that \approachNS~ relies on \textsc{EvoSuite} for assertion generation, which prevents it from exploiting the more semantically accurate assertions that LLMs can produce. It is also worth noting that neither \approach{}, nor \textsc{\cmj{}}, nor \textsc{EvoSuite} were configured for strong or weak mutation due to time constraints. Nonetheless, LLM-generated test snippets could effectively enhance assertion quality of EvoSuite.

\begin{rqbox}
\textbf{Answer to RQ1:}
\approachNS~ outperforms \textsc{\cmj{}}, \textsc{EvoSuite}, and \textsc{LLMOnly} in terms of median branch coverage, line coverage, and mutation score. However, the gains in mutation score are more modest compared to the improvements observed for branch and line coverage.
\end{rqbox}

\begin{table}[b!]
\centering
\caption{Effect size and significance of LLMSuite compared to the baselines (per metric).}
\begin{smaller}
\begin{tabular}{ll|cccc}
\toprule
\textbf{Comparison} &
\textbf{Metric} & \textbf{P-Value} & \textbf{Significant} & \textbf{A$_{12}$} & \textbf{95\% CI} \\
\midrule
\multirow{3}{*}{LLMSuite vs. \cmj{}}
  & Branch   & $1.6\times10^{-9}$  & Yes & 0.73 & [0.66, 0.78] \\
  & Line     & $3.9\times10^{-11}$  & Yes  & 0.74 & [0.68, 0.80] \\
  & Mutation & 0.0003                  & Yes  & 0.58   & [0.53, 0.65] \\
\midrule
\multirow{3}{*}{LLMSuite vs. EvoSuite}
  & Branch   & $5.5\times10^{-9}$  & Yes & 0.72 & [0.68, 0.80] \\
  & Line     & $3.7\times10^{-10}$  & Yes  & 0.80 & [0.60, 0.92] \\
  & Mutation & 0.007                  & Yes  & 0.57   & [0.50, 0.63] \\
\midrule
\multirow{3}{*}{LLMSuite vs. LLMOnly}
  & Branch   & $1.9\times10^{-21}$ & Yes & 0.85 & [0.79, 0.90] \\
  & Line     & $1.6\times10^{-29}$ & Yes  & 0.85   & [0.79, 0.90] \\
  & Mutation & 0.002                  & Yes  & 0.64   & [0.54, 0.74] \\
\bottomrule
\multicolumn{4}{l}{In terms of branch coverage:} \\
\multicolumn{4}{l}{No. cases LLMSuite better than \cmj{} (Lowest 95\% CI$>0.5$)} & \multicolumn{2}{r}{38 (38\%)}\\
\multicolumn{4}{l}{No. cases LLMSuite worse than \cmj{} (Highest 95\% CI$<0.5$)} & \multicolumn{2}{r}{3 (3\%)} \\
\multicolumn{4}{l}{No. cases LLMSuite better than LLMOnly (Lowest 95\% CI$>0.5$)} & \multicolumn{2}{r}{59 (59\%)}  \\
\multicolumn{4}{l}{No. cases LLMSuite worse than LLMOnly (Highest 95\% CI$<0.5$)} & \multicolumn{2}{r}{3 (3\%)}\\
\multicolumn{4}{l}{No. cases LLMSuite better than EvoSuite (Lowest 95\% CI$>0.5$)} & \multicolumn{2}{r}{36 (36\%)}\\
\multicolumn{4}{l}{No. cases LLMSuite worse than EvoSuite (Highest 95\% CI$<0.5$)} & \multicolumn{2}{r}{2 (2\%)} \\
\bottomrule
\end{tabular}
\end{smaller}
\label{tab:statistics}
\end{table}

\subsection{RQ2: Effect of Design Choices}
This RQ examines how different design decisions affect the effectiveness of our approach, measured in terms of branch coverage. Specifically, we analyze the impact of varying the LLM, using different search-based strategy, using different context levels (method vs. class), prompting strategy, and number of LLM calls. The \textsc{\cmj{}} baseline could also be considered a variant, as it uses a different search-based algorithm and prompting strategy. However, we exclude it from this RQ since it was already examined in detail in RQ$_1$.
Table~\ref{tab:branch_coverage_transposed} reports the mean and median coverage for each variant, along with their differences compared to \textsc{EvoSuite}.

\begin{figure*}[!ht]
\centering
\includegraphics[width=\linewidth]{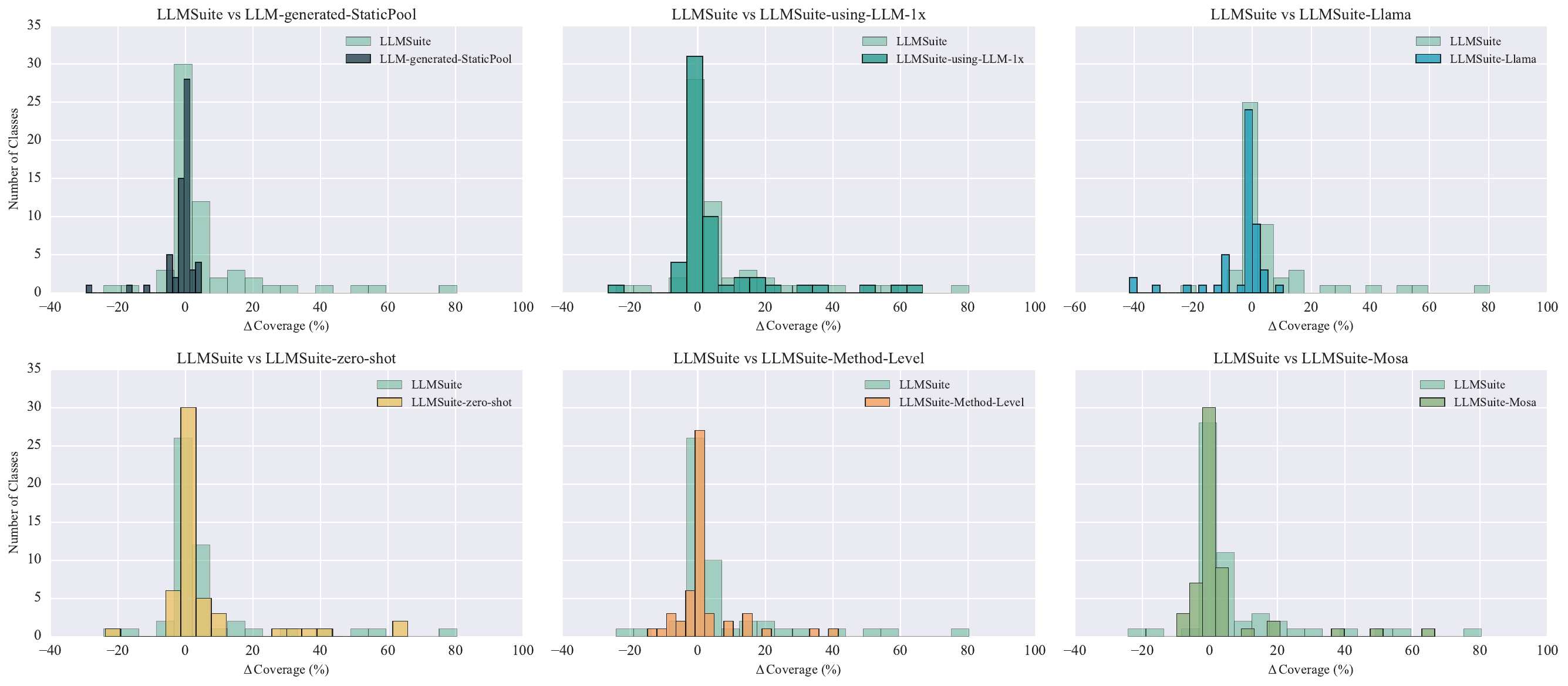}
\caption{Histogram of Per-Class Branch Coverage Improvement ($\Delta$) for Different \approachNS~ Variants Compared to \textsc{EvoSuite}}\label{fig:results_delta}
\vspace{-1mm}
\end{figure*}

To evaluate the contribution of each design choice, we compute the performance delta ($\Delta$) relative to \textsc{EvoSuite} as the baseline, and compare these results to the default configuration of \approachNS, introduced in the previous section. Figure~\ref{fig:results_delta} shows the distribution of $\Delta$-coverage across individual classes for each variant. Statistical analysis confirms that the default \approachNS~ configuration achieves significantly higher coverage than each variant ($p$-value$\ll$0.05), with a large effect size ($A_{12} > 0.71$) in all comparisons except against the Zero-Shot variant, where the effect size is smaller ($A_{12} = 0.60$).

\vspace{4pt}
\noindent\textbf{Effect of Different Context Levels}.
\revised{Unlike the default \textsc{LLMSuite}, which employs class-level prompting, this variant uses method-level prompting to assess the impact of context level (method vs.\ class) with the same underlying LLM (\textsc{ChatGPT}). On average, branch coverage improves by +3.76\% compared to \textsc{EvoSuite}, although the difference is not statistically significant ($p = 0.4$). However, the default class-level configuration of \textsc{LLMSuite} significantly outperforms the method-level variant ($p \ll 0.05$), highlighting the benefit of richer, class-level context.}

\begin{table}[b!]
\centering
\caption{Branch Coverage Metrics Across All Variants}
\begin{smaller}
\begin{tabular}{lcccccccc}
\toprule
\multirow{2}{*}{\textbf{Metric}} & \multirow{2}{*}{\textbf{EvoSuite}} & \multicolumn{7}{c}{\textbf{LLMSuite Variants}}\\
\cline{3-9}
 &  & \textbf{Orig.} & \textbf{S. Pool} & \textbf{1x} & \textbf{Llama} & \textbf{0-Shot} &  \textbf{Mosa} & \textbf{M. Level}\\
\midrule
Mean           & 56.48\% & 63.39\%  & 55.08\% & 61.85\% & 47.82\% & 61.29\% & 60.22\% & 60.23\% \\
Median         & 60.12\% & 74.81\% & 58.80\% & 70.00\% & 42.65\% & 67.67\% & 68.60\% & 62.53\% \\
$\Delta$ Mean         & --      & +6.91\% & -1.40\% & +5.37\% & -8.66\% & +4.82\% & +3.74\% & +3.76\% \\
$\Delta$ Median       & --      & +14.70\% & -1.32\% & +9.88\% & -17.46\% & +7.56\% & +8.49\% & +2.42\% \\
\bottomrule
\end{tabular}
\end{smaller}
\label{tab:branch_coverage_transposed}
\end{table}

\vspace{4pt}
\noindent\textbf{Effect of Using a Local LLM with Different Context Levels}.
\label{sec:result:llama}
We replaced ChatGPT with CodeLLaMA 13B to examine the effect of using a different LLM and different context level (method vs class). Due to the context-window limitations of CodeLLaMA, it was prompted at the method level instead of the class level, though the same parser was used. On average, branch coverage drops by 8.6\% compared to \textsc{EvoSuite}, with most classes showing lower coverage and only a few seeing slight improvements~(Figure~\ref{fig:results_delta}). Compared to ChatGPT, we hypothesize that CodeLLaMA is less effective at generating test cases that improve coverage. This may be due to the fact that the response time of CodeLLaMA is much slower than ChatGPT, leaving less time for the search process to explore the test space.

\vspace{4pt}
\noindent\textbf{Effect of Prompting Strategy}.
We evaluated a \textit{zero-shot prompting} variant, \textit{0-Shot}, to assess the impact of prompting strategy. Unlike the default \approach{}, which uses a self-refinement mechanism, this variant generates test cases in a single pass without iterative refinement. \textit{0-Shot} achieved a mean branch coverage improvement of $+4.82\%$ over the baseline (\textsc{EvoSuite}), with statistical significance ($p$-value=0.004). However, the default \approachNS~ still significantly outperforms the zero-shot variant ($p$-value=0.019), confirming the benefit of self-refinement in prompt generation.

\vspace{4pt}
\noindent\textbf{Effect of Reducing the Number of LLM-Test-Generation Calls}.
To assess the role of repeated LLM interactions, we compared the default \approachNS, which makes five LLM calls per target ($L=5$), with a simplified variant using only one call ($L=1$), denoted as \textit{1x}. This variant achieved a branch coverage improvement of $+5.37\%$ over \textsc{EvoSuite}, with statistical significance ($p$-value=0.002). 
These results show that even a single LLM call can yield significant coverage gains. However, multiple calls further improve performance ($p$-value$\ll$0.05), demonstrating the benefit of iterative prompting.

\vspace{4pt}
\noindent\textbf{Effect of Improving Only String Inputs}.
We examined whether enhancing only the string inputs affects coverage. \textsc{EvoSuite} generates strings using a static pool of random values, along with values collected through instrumentation. In this variant, we replaced the static pool with LLM-generated strings tailored to the class under test, without using LLM-generated tests to guide the search process.
The results show that this change alone does not significantly improve coverage; in fact, branch coverage dropped slightly by 1.4\% compared to \textsc{EvoSuite}. This suggests that improving string inputs in isolation is insufficient --- other aspects of the test generation process also need to be addressed.

\vspace{4pt}
\noindent\textbf{Effect of Using a Different SBST Strategy}.
\revised{Unlike the default \textsc{LLMSuite}, which uses DynaMOSA as its search algorithm together with the standard prompting and self-refinement strategy, this variant replaces DynaMOSA with MOSA to assess the impact of the underlying SBST algorithm. On average, branch coverage improves by +3.74\% compared to \textsc{EvoSuite}, although the difference is not statistically significant ($p = 0.3$). However, the default \textsc{LLMSuite} configuration significantly outperforms the MOSA-based variant ($p \ll 0.05$), suggesting that the stronger search strategy is better able to exploit the LLM-generated code snippets and translate them into additional coverage gains.}

\begin{rqbox}
\textbf{Answer to RQ2:}
  \revised{The default \approachNS~ configuration outperforms the 6 variants -- featuring a different LLM, a different context level, different prompting strategy, a reduced number of prompts to the LLM, only improving string inputs, and a different SBST strategy -- that we have investigated.} 
\end{rqbox}

\subsection{RQ3: Improvement over SBST  Approach}
\begin{figure}[!t]
\centering
\includegraphics[width=\linewidth]{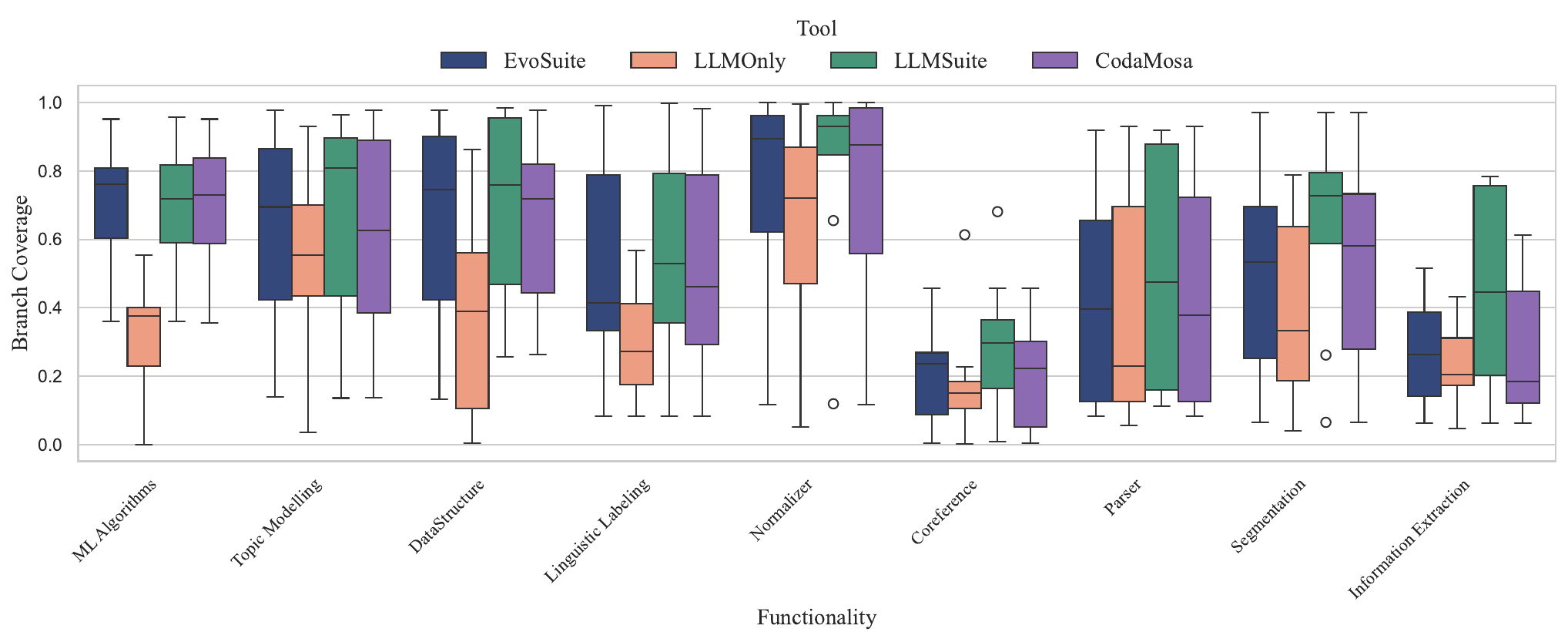}
\caption{Coverage Comparison per Different Functionality}\label{fig:results_category}
\end{figure}

We analyzed which components of NLP libraries are highly covered by \textsc{EvoSuite}-generated tests, LLM-based tests, and \approachNS. Figure~\ref{fig:results_category} shows that \approachNS~ consistently outperforms both baselines --- \textsc{EvoSuite} and LLM-Tests --- across all categories, except for Machine Learning, where it shows a marginal decline of 0.1\%.

The mean branch coverage improvements achieved by \approachNS~ across functionality categories are: Information Extraction (18\%), Coreference (8\%), Segmentation (7\%), Normalization (5\%), and Linguistic Labeling (3\%). The gains are especially prominent in classes that are text-heavy and involve language-specific processing, as opposed to those focused on numeric computation, data structures, or utility logic. To better understand this, we analyzed classes where the differences were most pronounced.

\vspace{2pt}
\noindent\textbf{How Using LLM can improve search-based test generation}. Our analysis reveals three main ways in which LLMs enhances coverage in \approachNS~ compared to traditional search-based methods:

\begin{listing}[b!]
\begin{minted}[fontsize=\footnotesize]{java}
// A. An invalid behavior in the EntityMentionsAnnotator class - CoreNLP Project 
// if anything is still at 1.1, set it to -1.0
for (String label : entityLabelProbVals.keySet()) {
  if (entityLabelProbVals.get(label) >= 1.1) {
    entityLabelProbVals.put(label, -1.0);
  }
}
// ----------------------------------------------------------------------------
// B. An LLM-generated test case that contributed to LLMSuite’s improved coverage.
@Test
public void testDetermineEntityMentionConfidences() {
  CoreLabel token = new CoreLabel();
  token.setWord("Barack");
  token.set(NamedEntityTagProbsAnnotation.class, Map.of("PERSON", 0.95, "LOCATION", 0.05));
  CoreMap entityMention = new ArrayCoreMap();
  entityMention.set(TokensAnnotation.class, List.of(token));
  EntityMentionsAnnotator.determineEntityMentionConfidences(entityMention);
  ...
}
\end{minted}
\caption{Example Scenario for EntityMentionsAnnotator}
\label{listing:entity}
\end{listing}

\textit{A. Test Data and sequence of object calls.}
In many test cases, LLMs provide useful guidance for constructing objects and invoking them in the correct order. This is particularly important in classes that require complex setup, where uninformed search may get stuck in local optima due to the size of the search space. For example, for three classes in the \textit{CoreNLP} project, in \texttt{QuoteAnnotator} (64\% improvement), \texttt{EntityMentionsAnnotator} (41\%), and \texttt{WordsToSentencesAnnotator} (39\%), the improvements mainly come from LLMs generating correct sequences of instantiations and method calls.

Listing~\ref{listing:entity} in item~\emph{B} shows a representative example for \texttt{EntityMentionsAnnotator}. The test constructs a \texttt{CoreLabel} token, assigns entity type probabilities (e.g., \texttt{PERSON} at 0.95), wraps it in an \texttt{ArrayCoreMap}, and invokes the target method. \textsc{EvoSuite} struggles with such setup due to the specificity and inter-dependencies among objects. Such guidance also enables the exploration of behavior-dependent branches, such as error handling for invalid probabilities, as illustrated in item A of Listing~\ref{listing:entity}.

\textit{B. Generating Domain-Specific Text Inputs.}
Many NLP classes rely on structured or linguistically rich inputs that random generation fails to trigger. We observe that LLM-generated test snippets enable \approachNS~ to better exercise parsing, segmentation, and normalization logic. For example, OpenNLP’s \texttt{Parser} class expects Treebank-style input structures. \approachNS~ was able to generate such input, as in the example below, which activates branches missed by \textsc{EvoSuite}:

\begin{minted}[fontsize=\footnotesize]{java}
Parse parse0 = Parse.parseParse("(TOP (NP (NN -LRB-) (NN book) (NN -RRB-)))");
\end{minted}

\textit{C. Inferring Configuration and Property Settings.}
Some classes rely on external properties or configuration to activate certain branches. For example, in \texttt{WordsToSentencesAnnotator}, certain HTML boundary handling logic is only triggered when the appropriate configuration is set, which is shown in Listing~\ref{listing:words}.

\begin{listing}[b!]
\begin{minted}[fontsize=\footnotesize]{java}
// A. LLMSuite-generated test case line 
properties0.setProperty("ssplit.htmlBoundariesToDiscard", "div,be");
// -----------------------------------------------------------------------------
// B. CoreNLP Project - WordsToSentencesAnnotator.java
// HTML boundaries which are discarded
bounds = properties.getProperty("ssplit.htmlBoundariesToDiscard");
if (bounds != null) {
    String[] elements = bounds.split(",");
    htmlElementsToDiscard = Generics.newHashSet(Arrays.asList(elements));
}
\end{minted}
\vspace{-3.5mm}
\caption{Example Scenario for WordsToSentencesAnnotator}
\label{listing:words}
\end{listing}

\vspace{2pt}
\noindent\textbf{Where LLM-Tests Provide Less Guidance}.
\textsc{LLMSuite} shows limited benefit in classes dominated by numeric computation or algorithmic logic, such as KMeans. In these scenarios, LLMs often fail to generate effective numeric inputs or encounter hallucinations, which we hypothesize can mislead the test generation process or trap it in local optima. As a result, the added value of LLM guidance is minimal, and in some cases, test quality may degrade slightly.

\begin{rqbox}
\textbf{Answer to RQ3:}
  \approachNS~ improves upon SBST when it comes to providing test data, getting the sequence of object calls right, and inferring configuration and property settings. \approachNS~ shows limited benefits in classes dominated by numeric computation and algorithmic logic. 
\end{rqbox}

\subsection{RQ4: Improvement over manual tests}\label{sec:results:rq4}

\begin{figure*}[!t]
\centering
\includegraphics[width=\linewidth]{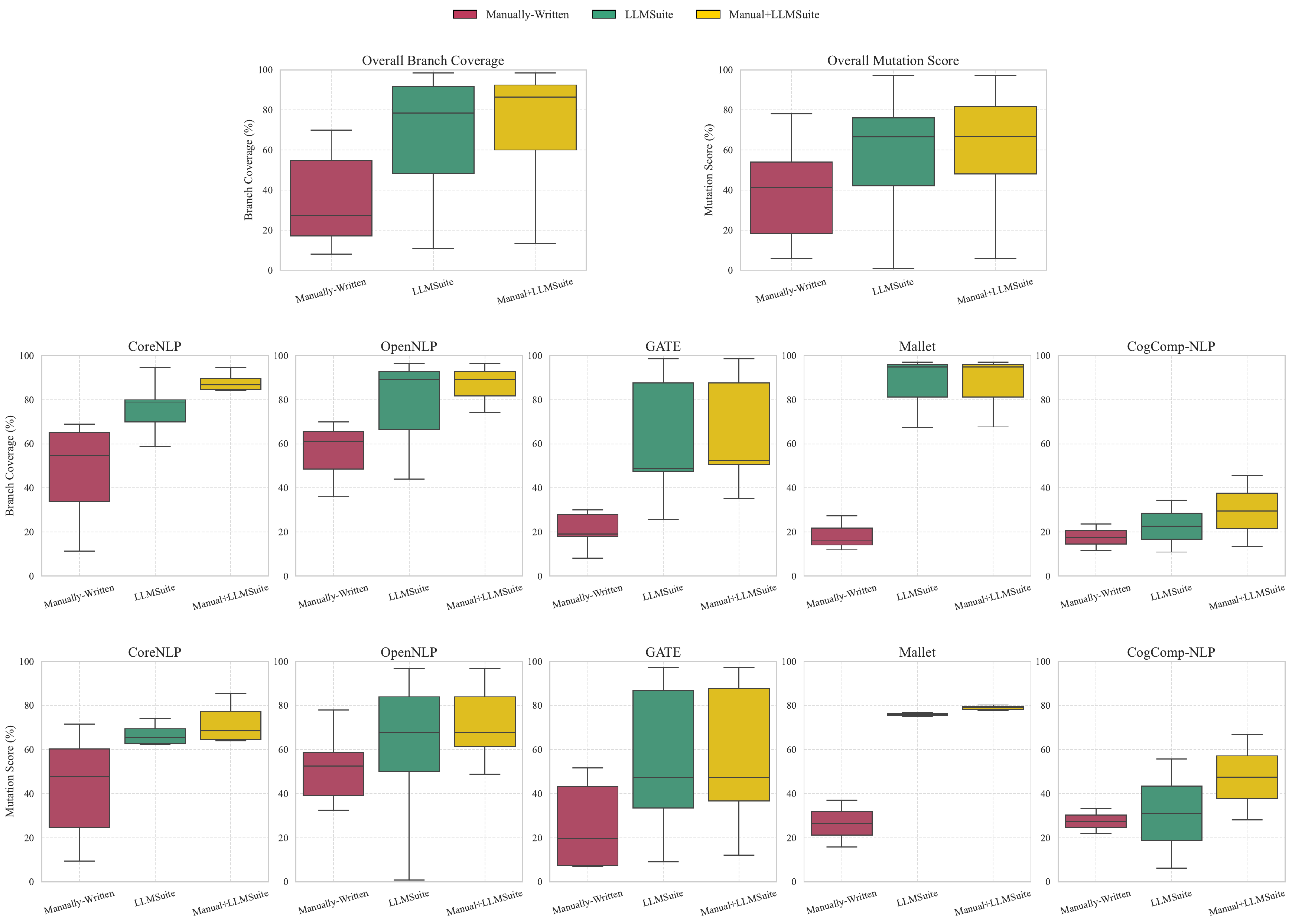
}
\caption{\revised{Comparison of Branch Coverage and Mutation Score Achieved by Manually Written and LLMSuite-Generated Test Cases Across %
Projects.}}\label{fig:results_manual}
\end{figure*}

Out of the 100 classes in our dataset, only 27 had dedicated manually written test cases; the remaining 73 may have been indirectly covered by other tests.  Manual tests in GATE use JUnit 3, those in OpenNLP use JUnit 5, and the three remaining projects 
--- CoreNLP, Mallet, and CogCompNLP --- 
primarily use JUnit 4. Because of this, we ran each project’s own setup with JaCoCo and \revised{PITest}, rather than through our unified coverage tool. \revised{We first examine how LLMSuite-generated tests compare with manually written tests in terms of branch coverage and mutation score, as well as how they complement the existing manual test suites. We then analyze the areas missed by manually written tests through a categorization of those gaps, from which we derive a taxonomy.}

\begin{table}[b!]
\centering
\caption{Comparison of branch coverage and mutation score across projects for manually written tests, LLMSuite-generated tests, and their combination}
\renewcommand{\tabcolsep}{4.5pt}
\begin{smaller}
\begin{tabular}{lc|cccc|cccc}
\toprule
\multirow{2}{*}{\textbf{Project}} & \multirow{2}{*}{\textbf{\#}} & 
\multicolumn{4}{c|}{\textbf{Branch Coverage}} &
\multicolumn{4}{c}{\textbf{Mutation Score}} \\
\cline{3-6}
\cline{7-10}
& & \textbf{Man} & \textbf{LLS} & \textbf{Man+LLS} & \textbf{$\Delta$}
& \textbf{Man} & \textbf{LLS} & \textbf{Man+LLS} & \textbf{$\Delta$} \\
\midrule
CoreNLP  & 7 & 47.53\% & 76.42\% & 85.60\% & +38.08\% & 43.02\% & 62.42\% & 67.37\% & +24.35\% \\
OpenNLP  & 7 & 55.67\% & 76.57\% & 86.63\% & +30.97\% & 47.63\% & 59.11\% & 65.37\% & +17.75\% \\
GATE     & 6 & 20.61\% & 61.69\% & 64.82\% & +44.21\% & 25.82\% & 54.79\% & 56.22\% & +30.40\% \\
Mallet   & 5 & 18.49\% & 86.47\% & 86.55\% & +68.06\% & 26.47\% & 76.04\% & 79.03\% & +52.56\% \\
CogCompNLP & 2 & 17.6\% & 22.65\% & 29.56\% & +11.96\% & 27.51\% & 31.03\% & 47.49\% & +19.98\% \\
\midrule
Overall  & 27  & 33.99\% & 68.49\% & 74.55\% & +40.55\% & 38.1\% & 58.06\% & 63.54\% & +25.44\% \\
\bottomrule
\end{tabular}
\end{smaller}
\label{tab:rq4_branch_mutation}
\end{table}

\subsubsection{Incremental Coverage and Mutation Score}

\revised{Figure~\ref{fig:results_manual} compares the branch coverage and mutation score achieved by manually written test cases and those generated by \approachNS, as shown in the box plots. As reported in Table~\ref{tab:rq4_branch_mutation}, manually written tests achieve a mean branch coverage of 33.99\%, whereas LLMSuite-generated tests achieve 68.49\%. When combined, the two test suites reach 74.55\%, corresponding to an improvement of 27.7\%.}

\revised{A similar pattern can be observed for mutation score. Manually written tests achieve 38.1\%, LLMSuite-generated tests achieve 58.06\%, and the combined suite reaches 63.54\%, corresponding to an improvement of 25.44\%. Overall, LLMSuite complements manually written tests in both branch coverage and mutation score, although its effect is more highlighted for branch coverage.}

\revised{This observation is consistent with the findings from RQ1, where we noted that LLMSuite relies on EvoSuite for assertion generation. As a result, it benefits less from LLM-generated assertions, which may explain why the improvement in mutation score is smaller than the improvement in branch coverage.}

\revised{Furthermore, the quality and completeness of manual tests varied across projects. Actively maintained ones, such as OpenNLP and CoreNLP, generally had more comprehensive test suites. In contrast, GATE, Mallet, and CogCompNLP exhibited lower coverage.}

\subsubsection{Taxonomy of Detected Gaps}
\revised{To investigate how \approachNS{} complements existing test coverage, we manually analyzed each project and its corresponding test cases to determine which parts of the production code were exercised by manually written tests and which areas remained untested. To better understand these uncovered regions, we manually reviewed the uncovered code blocks across the dataset and categorized the reasons for their lack of coverage. For this purpose, we adopted the classification scheme of Wang et al.~\cite{wang2021automatic}, originally proposed for analyzing test gaps in machine learning libraries. We adapted their scheme by adding two new categories---Method Overloading (MEO) and Properties-Dependent Behavior (PRB), marked with an asterisk (*)---and removing the message-handling behavior (MEB) category, which was not common in our dataset. This analysis showed that, in many cases, \approachNS{} was able to extend coverage to code blocks that were not reached by the manual test suites. In the following, we illustrate both the covered and uncovered behaviors using examples from the CoreNLP project.}

\vspace{3pt}
\noindent\textbf{What Manually-Written Tests Cover}. Manual tests tend to focus on a class’ main functionality through high-level scenarios, often combining multiple use cases in a single test method.
For example, the \code{CleanXmlAnnotator} class processes XML input and extracts information from tags like \texttt{<post>}, \texttt{<quote>}, and self-closing tags such as \texttt{<img/>}. These tags may include attributes like \texttt{author}, \texttt{datetime}, or \texttt{document\_id}. A sample XML snippet is shown in Listing~\ref{listing:cleanxml}.
The manual tests for this class cover some valid cases, such as well-formed posts and quotes with attributes, and a couple of invalid ones, like an unclosed tag. While these are meaningful scenarios, they do not cover the full range of possible behaviors.

\begin{listing}[!b]
\begin{minted}[fontsize=\footnotesize]{xml}
<post author="UDDep" datetime="2010-05-30T15:43:00" id="p2">
  <quote orig_author="James Rood">
    Yesterday afternoon as I negotiated route 149 from Lake George to Fort Ann
    in NY I passed a new diner that had opened that day.
    <img src="http://britishexpats.com/forum/images/smilies/wink.gif"/>
  </quote>
  If they don't have english food and beer...tell em...
</post>
\end{minted}
\vspace{-3mm}
\caption{An Example Scenario for CleanXMLAnnotator}
\label{listing:cleanxml}
\end{listing}

\vspace{3pt}
\noindent\textbf{What Manually-Written Tests Do Not Cover}. In contrast, manually written tests often omit alternative execution paths, special input configurations, and less central behaviors. In the following, we describe each gap category using examples from the CoreNLP project:

\textit{Valid Behaviors (VB).} A class often supports multiple valid behaviors, but only a subset is exercised by manual tests. For instance, in \texttt{CleanXMLAnnotator}, the tests check whether an XML document starts and ends with a specific tag and whether tags include attributes. However, they cover only a few specific tags, leaving many supported ones untested. This partial coverage explains why some lines remain uncovered.
Across all classes, 530 code blocks were left uncovered for this reason, accounting for 49\% of all uncovered blocks. With \approach{}, only 222 such blocks remained, primarily due to challenges in initializing complex objects.

\textit{Invalid Behaviors (IVB).}
In addition to valid behaviors, a class may also exhibit invalid behaviors when given inputs outside its expected domain, e.g., supplying a parameter with values beyond its allowable range.
While manually written tests cover some cases (e.g., unclosed XML tags in Listing~\ref{listing:cleanxml}), they often miss others, such as having more closing than opening tags. This leads to 56 coverage drops across the manually written suite. In contrast, \approach{} reduces this to just 16 cases by better addressing such edge scenarios.

\textit{Exception Handling (EX).} Manually written tests often do not cover exceptions. For instance, in \texttt{CleanXMLAnnotator}, an 
exception is thrown when the appropriate tag annotation is missing, yet this is untested. We have discovered 56 such exception cases in the manual suite, whereas \approach{} reduces this number by 11.

\textit{Auxiliary Methods (AUX).} Our analysis reveals that many utility or helper methods remain untested in manually written tests, primarily because they are declared as private and are not directly invoked in typical test scenarios. This  leaves 266 blocks uncovered, approximately 25\% of all uncovered code. While \textsc{LLMSuite} and search-based techniques can sometimes reach these methods indirectly, testing private code remains challenging because bytecode instrumentation does not provide  information about them.

\textit{Method Overloading (MEO).} Constructors and methods can have multiple overloaded variants, each with different parameter lists. In our dataset, 121 code blocks remained uncovered because not all overloads were exercised by manually written tests. The search-based process in \approach{} reduces this number to just 7 blocks.

\textit{Properties-Dependent Behavior (PRB).}
Some functionalities in NLP libraries depend on specific properties --- such as configuration files, model paths, or input documents --- to activate certain code paths. For instance, in \texttt{CleanXMLAnnotator}, different property values trigger different execution branches. These scenarios are often missed in manually written tests, leaving 48 blocks uncovered. \approach{} reduces this to 10, although handling such cases remains challenging due to the domain knowledge required.

\begin{rqbox}
\textbf{Answer to RQ4:}
  \revised{Combining manually written and \approachNS~generated test cases increases coverage and mutation score significantly}. We have identified 6 specific situations where \approachNS~ was able to make a strong contribution to additional test coverage.
\end{rqbox}

\begin{table}[!hb]
\caption{Distribution of uncovered code in each NLP library}
\label{tab:rq4_compact}
\centering
\begin{smaller}
\begin{tabular}{llrrrrrr}
\toprule
\textbf{Project} & \textbf{Tool} & \textbf{VB (\%)} & \textbf{IVB (\%)} & \textbf{EX (\%)} & \textbf{AUX (\%)} & \textbf{MEO (\%)} & \textbf{PRB (\%)} \\
\midrule
\multirow{2}{*}{CogComp-NLP} & LLS & 3 (18) & 1 (6) & 0 (0) & 0 (0) & 0 (0) & 13 (76) \\
                    & Man & 6 (46) & 3 (23) & 1 (8) & 0 (0) & 0 (0) & 3 (23) \\
\multirow{2}{*}{CoreNLP} & LLS & 33 (62) & 3 (6) & 5 (9) & 3 (6) & 0 (0) & 9 (17) \\
                    & Man & 41 (30) & 9 (7) & 10 (7) & 16 (12) & 31 (23) & 30 (22) \\
\multirow{2}{*}{GateCore} & LLS & 114 (33) & 10 (3) & 23 (7) & 179 (53) & 3 (1) & 12 (4) \\
                    & Man & 261 (39) & 36 (5) & 30 (4) & 209 (31) & 48 (7) & 8 (1) \\
\multirow{2}{*}{Mallet} & LLS & 15 (54) & 0 (0) & 2 (7) & 10 (36) & 0 (0) & 1 (4) \\
                    & Man & 154 (70) & 6 (3) & 7 (3) & 27 (12) & 40 (18) & 0 (0) \\
\multirow{2}{*}{OpenNLP} & LLS & 57 (60) & 2 (2) & 3 (3) & 23 (24) & 4 (4) & 4 (4) \\
                    & Man & 68 (68) & 2 (2) & 6 (6) & 14 (14) & 2 (2) & 7 (7) \\
\midrule
\multicolumn{2}{l}{\textbf{Overall LLS}} & 222 (42) & 16 (3) & 33 (6) & 215 (40) & 7 (1) & 38 (7) \\
\multicolumn{2}{l}{\textbf{Overall Man}} & 530 (49) & 56 (5) & 54 (5) & 266 (25) & 121 (11) & 48 (4) \\
\bottomrule
\end{tabular}
\end{smaller}

\end{table}

\section{Discussion}

\noindent\textbf{Test-generation time budget.}
We allocated a 15-minute time budget per target class. Within that time budget, LLM-based test generation with self-refinement on average consumes 148.2 seconds (16.1 seconds in zero-shot mode). This can reduce the time available for the search process when time is limited. However, since LLM test generation is independent of the search, this overhead can be avoided by running it in parallel with the SBST process. 

\vspace{2pt}

\noindent\textbf{Cost Analysis and Token Usage.}
\begin{figure*}[!t]
\centering
\includegraphics[width=\linewidth]{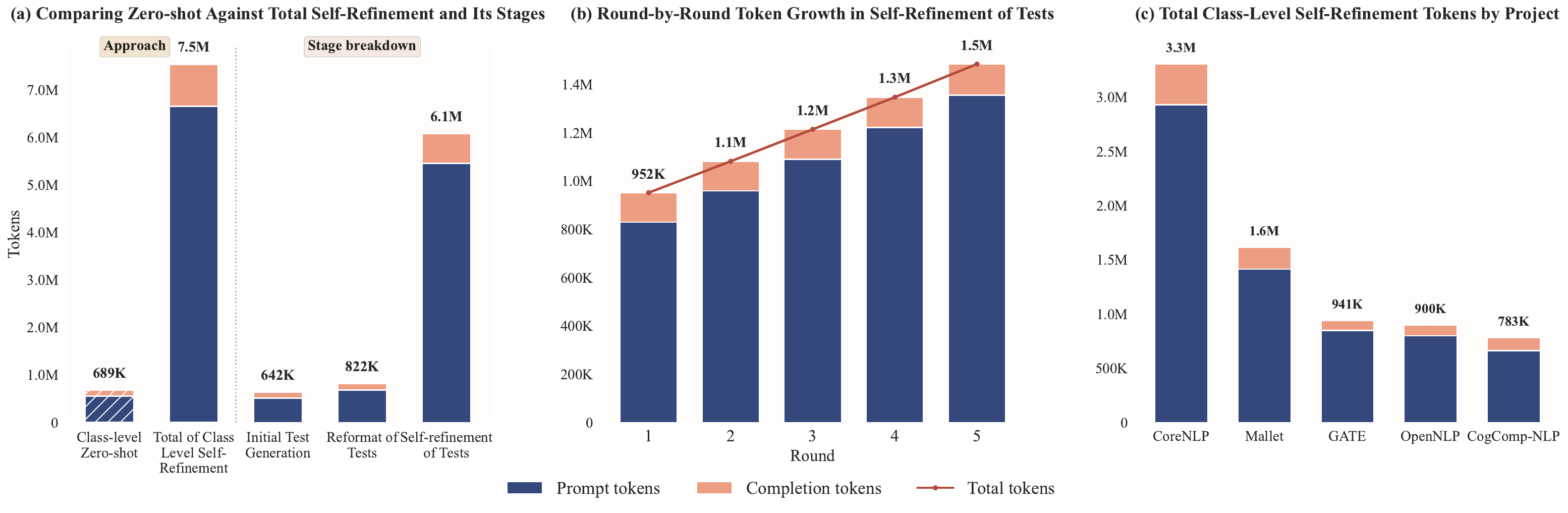}
\caption{Token usage analysis for class-level self-refinement on the 65 classes common to both class-level zero-shot and total class-level self-refinement.}
\label{fig:self_refinement_tokens}
\end{figure*}
\revised{Beyond increasing processing time, the practical cost of LLM-based test generation must also be considered. We analyzed token usage on the 65 common classes, out of the 100 selected classes in our experiment, for which results were available for both class-level zero-shot generation and total class-level self-refinement. As shown in Figure~\ref{fig:self_refinement_tokens}(a), the full self-refinement pipeline consumed 7,547,600 tokens over 455 LLM calls, whereas class-level zero-shot required 688,549 tokens over 65 calls. Assuming a pricing model of \$5.00 per one million input tokens and \$15.00 per one million output tokens, this corresponds to an estimated total cost of \$46.62 for self-refinement versus \$4.78 for zero-shot, or about \$0.72 versus \$0.07 per target class.}

\revised{Most of this overhead comes from self-refinement itself. Figure~\ref{fig:self_refinement_tokens}(a) shows that, out of the total 7,547,600 tokens, 6,082,787 (80.6\%) were consumed by the self-refinement rounds, while initial test generation and test reformatting accounted for 642,430 (8.5\%) and 822,383 (10.9\%), respectively. Figure~\ref{fig:self_refinement_tokens}(b) further shows that token usage grows steadily across rounds, from 952,357 in Round~1 to 1,485,163 in Round~5. This is consistent with prompt growth when chat history is retained. For example, in the \texttt{RelationFeatureExtractor} class from CoreNLP, the number of input tokens increased from 1572 to 3325, 5182, 6996, and 8684 across refinement iterations, while the output size remained relatively stable at around $\sim$1727 tokens.}

\revised{Figure~\ref{fig:self_refinement_tokens}(c) also shows variation across projects. The token usage correspond to 27 target classes for CoreNLP, 14 for Mallet, 7 for GATE, 8 for OpenNLP, and 9 for CogComp-NLP. After normalization, the average token usage per class is 122,457 for CoreNLP, 115,475 for Mallet, 134,482 for GATE, 112,549 for OpenNLP, and 86,984 for CogComp-NLP, suggesting that token cost also depends on class complexity and prompt size.}

\revised{Overall, these results show that the additional gains of iterative refinement come at substantial token and monetary cost. Therefore, the choice of the number of refinement iterations ($T$) should be guided not only by effectiveness gains but also by cost-effectiveness. In practice, a smaller number of refinement rounds may offer a better trade-off than using the full iterative process.}

\begin{figure*}[!t]
\centering
\includegraphics[width=0.7\linewidth]{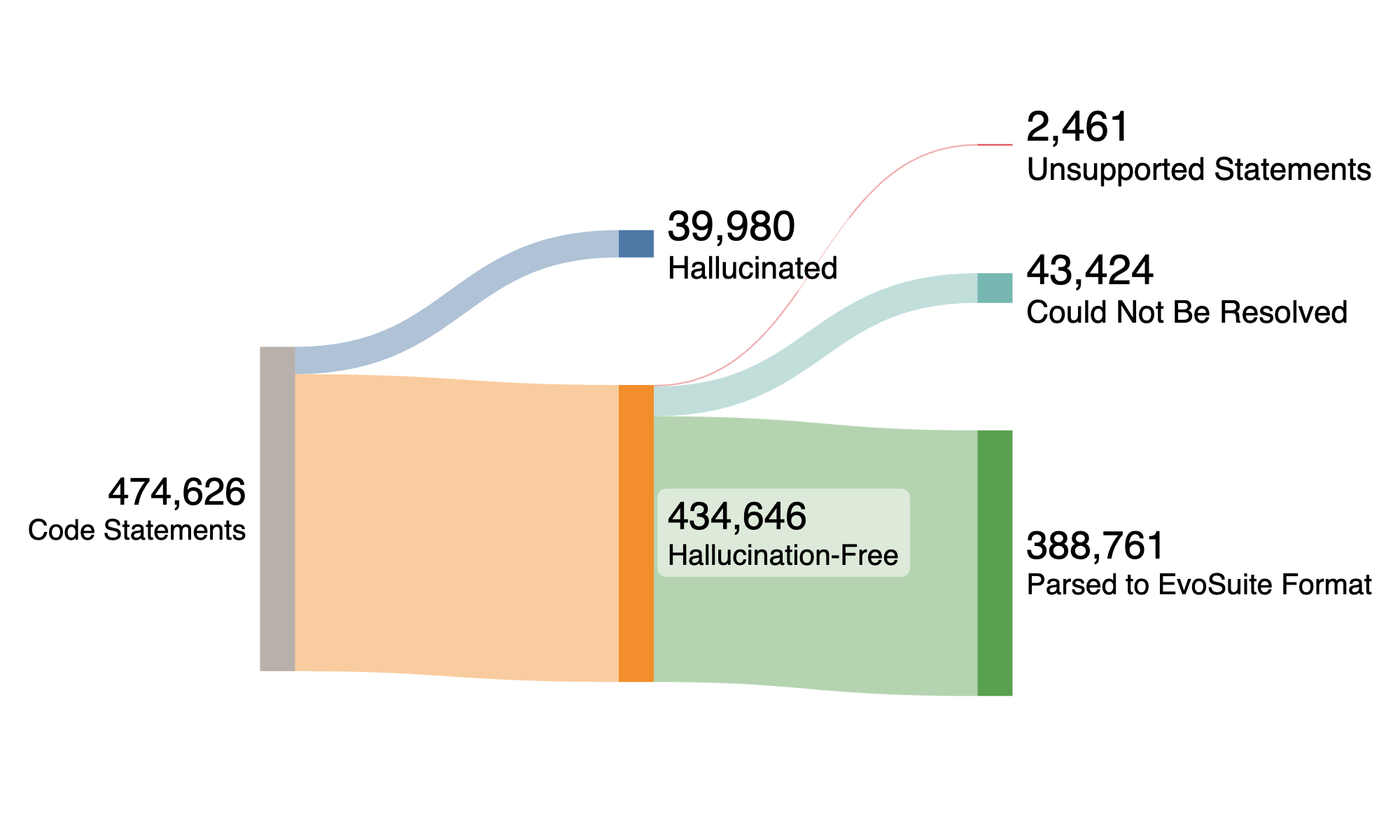}

\caption{Flow of LLM-generated statements through the cleaning and parsing pipeline.} \label{fig:parser}
\end{figure*}

\vspace{2pt}
\noindent\textbf{Parser.}
\revised{As discussed in Section~\ref{sec:approach}, our parser, \emph{LLM2EvoSuiteParser}, is able to handle complex cases such as nested method invocations, varargs, and multi-dimensional arrays. Still, limitations in both EvoSuite and Spoon prevent \approachNS~ from fully utilizing all LLM-generated test snippets. Figure~\ref{fig:parser} summarizes the flow of LLM-generated statements through the hallucination-filtering and parsing stages and shows how they are filtered into different outcomes.}

\revised{Before parsing, the generated code goes through a hallucination-filtering stage. In the LLM component, we first try to repair generated snippets using the heuristics described earlier. Because using an LLM-in-the-loop repair mechanism would be too computationally expensive, we rely only on heuristic-based fixes and try to preserve as much of the generated code as possible. Across 474,626 LLM-generated statements, we found 39,980 hallucinated statements, including API mismatches. Rather than discarding the full test case, we comment out the hallucinated lines together with any dependent lines that become invalid as a result, while preserving the remaining valid statements so they can still contribute to testing. This leaves 434,646 hallucination-free statements and compilable, meaning that 91.57\% of the generated lines remain usable. These cleaned tests are treated as \textit{LLMOnly} test cases, for which we measure coverage independently, and are then passed to \emph{LLM2EvoSuiteParser} for conversion into EvoSuite-compatible format.}

\revised{Out of the 434,646 hallucination-free statements, 388,761 were successfully converted, corresponding to a conversion success rate of 89.44\%. Among the remaining statements, 2,461 could not be parsed because they contain constructs that EvoSuite does not support, including \texttt{for} loops (691), \texttt{foreach} loops (605), \texttt{while} loops (515), \texttt{if} statements (344), and 306 other unsupported constructs such as new implementations of classes or interfaces. In addition, 43,424 statements could not be resolved during parsing. Most of these (40,393) involve API calls that EvoSuite does not support in the test generation process, such as \texttt{when()}, \texttt{spy()}, and \texttt{fail()}. The remaining 3,031 statements include unsupported binary or unary expressions, such as mathematical operations or string manipulations like \texttt{append()}.}

\revised{EvoSuite also inserts additional constructs --- such as catch blocks, assertions, and mock parameter initializations --- after the search phase, which prevents some LLM-generated lines from being translated directly. Spoon, on the other hand, sometimes fails to resolve types that are not explicitly defined. To address this, we adopt loose type-checking, which covers most of these cases. We also observed issues in EvoSuite’s conversion of test cases to source code when handling parameterized types with dependent generics, for example \texttt{class Value<J, T extends Type>}, where the second type depends on the first.}

\vspace{2pt}
\noindent\textbf{LLM Hallucination.}
\revised{LLMs are known to hallucinate in their outputs~\cite{yao2023llm}. In the context of code generation, this typically manifests as references to non-existent APIs~\cite{eghbali2024hallucinator}. As part of the hallucination-filtering stage, we first attempt to repair such issues using heuristics; if repair is not possible, the hallucinated lines are filtered out. Across all runs, we observed 39,980 hallucinated lines out of 474,626 generated lines (figure~\ref{fig:parser}).}

\revised{To better understand the nature of these hallucinations, we conducted a detailed analysis on generated test cases for 70 target classes in one run. In this subset, we identified \textit{8,449} hallucinated lines.}
We grouped these hallucinations into five categories: \textit{Unresolved References} (3,087 instances, $\sim$36.5\%), \textit{Access Modifier Issues} (1,621, $\sim$19.2\%), \textit{Abstract Instantiation Errors} (1,573, $\sim$18.6\%), \textit{Type Mismatches or Incompatible Types} (1,177, $\sim$13.9\%), and \textit{Other} (991, $\sim$11.7\%).
We also investigated whether there is any correlation between hallucination frequency and coverage improvement: 
we computed the Pearson correlation coefficient between $\Delta$Coverage and the ratio of commented hallucinated lines to the number of executed tests ($\Delta~\text{Coverage}~ \sim~\frac{\#\text{Errors Commented}}{\#\text{Tests Executed}}$). The result was \emph{–0.273}, which indicates a weak negative correlation. This suggests classes with more hallucinations per test tend to benefit less from \approachNS~ in terms of coverage improvement.

\vspace{2pt}
\noindent\textbf{Training Contamination.}
ChatGPT-4o-latest-2025-07-12 has been trained on a large corpus of publicly available code, including GitHub. Since the exact  training data has not been disclosed by OpenAI, it is not possible to confirm whether any code from our dataset was included. To assess potential overlap, we use JPlag~\cite{prechelt2002finding}, a code similarity detection tool, to compare LLM-generated test cases with manually written ones. The results indicate that the average of the average similarity is 4.48\%, and the average of the maximum similarity is 12.05\%, both relatively low scores. 
Additionally, as discussed in Section~\ref{sec:results:rq4}, only 30 classes in our dataset have corresponding manually written test cases. 
Among the few cases with higher similarity scores, we selected the top five for manual inspection. This analysis revealed no meaningful overlap, indicating that ChatGPT is not replicating the manually written test cases.

\vspace{2pt}
\noindent\textbf{Practical Value and Scope of LLMSuite.}
\revised{Although \approachNS~ improves structural coverage, these gains should be interpreted carefully. Search-based unit testing techniques such as EvoSuite are intended for structural exploration rather than specification-based validation~\cite{Fraser_2013,TOSEM_evaluation}. In particular, regression-based oracles capture observed behavior, not necessarily intended functionality, and therefore do not guarantee semantic correctness. The generated tests should thus be viewed as complementary to manually written or specification-based tests.}

\revised{Coverage is also only a proxy for test effectiveness~\cite{inozemtseva2014coverage}. While higher coverage increases the chance of exercising diverse behaviors, it does not necessarily imply stronger fault detection. For this reason, we complement coverage with mutation score, which provides complementary evidence of fault-detection capability~\cite{just2014mutants,Shamshiri_2015,andrews2005mutation,papadakis2018mutation}.}

\revised{We also acknowledge that many industrial testing challenges arise at higher levels, such as integration, system, and acceptance testing~\cite{ArcuriEMSE2018,10.1145/3680463,arcuri2018evomaster}. Still, automated unit test generation remains useful for complex libraries, where manually exploring edge cases is difficult. In this work, RQ3 and RQ4 examine the practical value of the approach beyond coverage alone by showing how \approachNS~ complements both SBST and manually written tests through exercising additional code regions and behaviors. These results suggest that the main benefit of the approach lies not only in higher coverage, but also in broader behavioral exploration, especially in cases where existing techniques struggle.}

\section{Threats To Validity}
We identify the following threats to the validity of our results:

\noindent\textbf{Internal Validity.}
Threats may arise from errors in the LLMSuite implementation or the evaluation pipeline. While some limitations were discussed earlier, we manually verified key components, wrote test cases to validate correctness, and inspected portions of the results. Additionally, we cross-validated coverage and mutation scores by rerunning experiments and reviewing outliers. %

\vspace{2pt}
\noindent\textbf{External Validity.}
We evaluated 100 classes from five open-source NLP libraries, systematically selected to cover diverse functionalities. %
\revised{While this provides a diverse and challenging sample, it is not intended to be representative of all classes in NLP libraries. Instead, our benchmark focuses on structurally complex, hard-to-cover units. This design choice is intentional, as such non-trivial classes are typically more difficult and costly to test manually, making them a more suitable target for evaluating the practical benefits of automatic test generation~\cite{PANICHELLA2018236}.} The results may also not generalize to other software domains, such as machine learning. Additionally, while we used LLMs trained on open-source code (GPT-4o, CodeLLaMA-13B), our similarity analysis revealed no meaningful overlap with manually written tests, indicating training data leakage~\cite{sallou2024breaking} did not affect our results.

\vspace{2pt}
\noindent\textbf{Construct Validity.}
While we evaluate test effectiveness using standard metrics such as line/branch coverage and mutation score, they do not fully capture qualitative aspects like test readability or maintainability, or the ability to detect real-world faults. Moreover, although we implemented \cmj{} as a Java counterpart to the original CodaMosa (as described in Section~\ref{sec:experiment}), it may not perfectly replicate exactly the original framework, introducing a potential threat to validity.

\vspace{2pt}
\noindent\textbf{Conclusion Validity.}
To ensure the reliability of our comparisons, we followed established guidelines for evaluating search-based algorithms~\cite{arcuri2014hitchhiker} and studies involving LLMs~\cite{sallou2024breaking}. We used appropriate statistical tests, such as the Wilcoxon signed-rank test, along with effect size measures. However, variability in LLM outputs and the inherent non-determinism of search-based methods can introduce noise into the results. To mitigate this, we averaged results over multiple runs and analyzed confidence intervals where applicable.

\section{Related Work}

\subsection{Test Cases for NLP Libraries}

As discussed earlier, testing NLP and ML libraries comes with its own set of challenges that differ from traditional software systems. The complexity of the data and the wide range of possible scenarios make testing these libraries particularly difficult. As a result, their unit tests often show lower coverage and mutation scores, and important aspects such as bias, fairness, and security are not consistently evaluated~\cite{10.1145/3680463, wang2021automatic}.

Researchers have tried to address these issues in different ways. For example, NLPLego~\cite{10.1145/3691631} improves metamorphic testing by checking whether NLP models produce the expected outputs under structured input transformations. It helps uncover subtle inconsistencies by generating diverse and valid inputs.
To improve unit-level test generation, other studies~\cite{mitchell, krodinger2025constraint} have explored combining grammar-based fuzzing with SBST. Some focus on handling structured inputs like XML and JSON~\cite{mitchell}, while others generate type-consistent inputs for complex ML frameworks such as TensorFlow and PyTorch~\cite{krodinger2025constraint}. However, these methods are usually limited to narrow domains and depend on handcrafted grammars or custom type specifications, which reduces their scalability and general applicability.

LLMSuite addresses these limitations by using LLMs to generate complex and diverse test inputs, including edge-case scenarios, at the unit-test level. Instead of targeting a single data format, LLMSuite adapts to a variety of input structures. While the approach is generalizable to other domains, this study focuses on NLP libraries due to their text-heavy nature and the unique challenges they pose for automated testing.

\subsection{LLM-based Test Case Generation}

There are several LLM-only test generators, ranging from simple tools--such as the approach by Siddiq et al.~\cite{siddiq2024using}, which relies solely on LLMs to produce tests--to slightly more advanced ones like ChatTester~\cite{10.1145/3660783}, which introduces a basic compiler-error repair loop. More sophisticated systems such as TestSpark~\cite{sapozhnikov2024testspark}, CoverUp~\cite{pizzorno2024coverup}, and ChatUniTest~\cite{chen2024chatunitest} enrich prompts or iteratively fix errors to improve test quality. However, these approaches rely exclusively on LLMs, do not integrate SBST, and none of them focus on domain-specific software such as NLP libraries. Prior work also shows wide variation in outcomes: Siddiq et al. report only 2\% coverage on the SF110 dataset, whereas TestPilot~\cite{10329992} achieves around 70\% statement coverage on small JavaScript systems, and other studies aim to improve human-written tests. A few hybrid approaches--such as UTGen~\cite{deljouyiICSE2025} and Aster~\cite{11121720}--combine LLMs with SBST, but mainly to improve understandability and readability.

\revised{More recent work combines LLMs with complementary guidance mechanisms rather than relying on them alone. Yang et al.~\cite{yang2024advancing} propose a program-analysis-guided approach that uses dependency information and counter-examples to steer the LLM toward hard-to-cover branches, reporting strong coverage improvements over prior SBST and LLM-based baselines. Similarly, Broide and Stern~\cite{broide2025evogpt} propose EvoGPT, a hybrid method that uses LLMs to generate diverse initial seeds and then applies evolutionary search to improve coverage and mutation score. These studies further support the view that LLMs are most effective when integrated with complementary techniques. LLMSuite follows the same general direction, but differs in its focus on NLP libraries, domain-specific prompting, self-refinement, and the conversion of LLM-generated code into EvoSuite-compatible test components that can be exploited during search.}

One notable hybrid LLM-guided SBST approach is \textsc{CodaMosa}~\cite{lemieux2023icse}, which also uses LLMs to escape local optima during search. While our approach shares this motivation, it introduces a self-refinement prompting strategy at the class level, whereas CodaMosa operates at the method level. In addition, we employ DynaMosa~\cite{Panichella2018AutomatedTC}, which has been shown to outperform MOSA~\cite{panichella2015reformulating} in prior studies~\cite{Panichella2018AutomatedTC, CAMPOS2018207, lukasczyk2023empirical}. CodaMosa also targets general-purpose classes rather than domain-specific software such as NLP libraries. Since no Java implementation of CodaMosa was available, we re-implemented it based on the published design. As shown in Section~\ref{sec:rq1}, LLMSuite achieves 11\% higher branch coverage, 8\% higher line coverage, and 11\% higher mutation score on NLP components.

\section{Conclusion}
Automated unit test generation remains particularly challenging for domain-specific software, such as NLP libraries, where test inputs must satisfy semantic (e.g., domain-specific knowledge), syntactic, and structural (e.g., input data format) constraints. In this paper, we proposed \textsc{LLMSuite}, a hybrid framework that integrates Large Language Models (in particular \texttt{ChatGPT-4o-latest-2025-07-12}) into the search-based test generation approach within  EvoSuite. When no search objective improves over multiple generations (search stagnation), \textsc{LLMSuite} invokes LLMs to generate class-level test suites with a self-refinement prompt strategy based on semantics (e.g., expressed in the JavaDoc and comments) and the structure of the class under test. %
These LLM-generated tests are injected into EvoSuite's evolving population.

Across our evaluation of 100 classes from five widely used Java NLP libraries, \textsc{LLMSuite} improves branch and line coverage by 15\% and mutation score by 5\% over EvoSuite, outperforms an LLM-only baseline by 36\% in structural coverage and approximately 24.7 percentage points in mutation score, and achieves 10\% and 8\% higher branch and line coverage than our Java re-implementation of \textsc{CodaMOSA}, with an 11\% gain in mutation score (RQ1). When investigating design alternatives (RQ2), we found that the default configuration of \textsc{LLMSuite} consistently outperforms variants using different LLMs, alternative prompting strategies, fewer prompts, or string-only improvements--demonstrating the importance of class-level prompting, self-refinement, and tight integration with SBST. Our qualitative analysis (RQ3) shows that \textsc{LLMSuite} is particularly effective when meaningful inputs, correct API invocation sequences, or domain-specific configuration settings are required, while offering limited advantages for classes dominated by numerical computation or algorithmic logic. Finally, in examining its interaction with manually written tests (RQ4), we found that \textsc{LLMSuite} complements manual test suites by exercising behaviors and configurations that those tests do not reach, and combining the two leads to additional coverage through several recurring forms of complementary behavior.

    \revised{We identify several directions for future work. First, an important extension is to compare \textsc{LLMSuite} against agent-based LLM systems that use compiler or execution feedback for iterative self-repair. While such approaches are promising, our framework was designed to give us full control over the entire generation, parsing, repair, and search pipeline within EvoSuite. Nevertheless, a direct comparison with external agentic baselines would be valuable future work. Second, future work should evaluate \textsc{LLMSuite} with other open-weight and closed-source LLMs. In particular, we plan to investigate the impact of other general-purpose models with different capabilities and alignment strategies. Because the framework is model-agnostic, these models can be incorporated without changing the overall architecture, enabling a broader assessment of generality across the evolving LLM landscape. We also plan to experiment with code-specific LLMs, e.g., StarCoder and DeepSeek Coder. In addition, it would be interesting to study whether integrating LLM-generated assertions more directly into the pipeline can improve behavioral validation beyond EvoSuite’s regression assertions. Finally, we plan to investigate \textsc{LLMSuite} in other domains where structured inputs and domain-specific semantics pose challenges for test generation, such as scientific computing libraries, parsers, and compilers.}

\begin{acks}
 This research was partially funded by the Dutch science foundation NWO through the Vici ``TestShift'' grant (No. VI.C.182.032). %
\end{acks}

\bibliographystyle{ACM-Reference-Format-num}
\bibliography{main}
\end{document}